 \documentclass[final,3p,times]{elsarticle}

 \usepackage{graphicx}
\usepackage{amssymb}
\usepackage{url}
\usepackage{grffile}

\begin{document}

\begin{frontmatter}



\title{A novel graph-based formulation for characterizing morphology with application to organic solar cells}


\author[ME]{Olga Wodo}
\author[ECpE]{Srikanta Tirthapura}
\author[ECpE,MSD]{Sumit Chaudhary}
\author[ME,ECpE]{Baskar Ganapathysubramanian\corref{cor}}
\address[ME]{Department of Mechanical Engineering, Iowa State University, Ames, Iowa,}
\address[ECpE]{Department of Electrical and Computer Engineering, Iowa State University, Ames, Iowa}
\address[MSD]{Department of Materials Science and Engineering, Iowa State University, Ames, Iowa}
\cortext[cor]{Tel.:+1 515-294-7442; fax: +1 515-294-3261. E-mail address: baskarg@iastate.edu, URL: http://www3.me.iastate.edu/bglab/ }

\begin{abstract}
Organic solar cells (OSC) have the potential for widespread usage, due to their promise of low cost, roll-to-roll manufacturability, and mechanical flexibility. However, ubiquitous deployment is impeded by their relatively low power conversion efficiencies (PCE). The last decade has seen significant progress in enhancing the PCE of these devices through various strategies. One such approach is based on morphology control. This is because morphology affects all phenomena involved in solar conversion: (1)~light absorption and electron-hole pair (exciton) generation; (2)~exciton diffusion and dissociation into free charges; and (3)~transport of charges to the electrodes.

Progress in experimental characterization and computational modeling now allow reconstruction and imaging of thin film morphology.
This opens up the possibility of rationally linking fabrication processes with morphology, as well as morphology with performance. In this context, a comprehensive set of computational tools to rapidly quantify and classify the 2D/3D heterogeneous internal structure of thin films will be invaluable in linking process, structure, and property.

We present a novel graph-based framework to efficiently compute a broad suite of physically meaningful morphology descriptors. 
These morphology descriptors are further classified according to the physical subprocesses within OSCs --- photon absorption, exciton diffusion, charge separation, and charge transport.  This approach is motivated by the equivalence between a discretized 2D/3D morphology and a {\it labeled, weighted, undirected graph}. We utilize this approach to pose six key questions related to structure characterization. We subsequently construct estimates and rigorous upper bounds of various efficiencies. The approach is showcased by characterizing the effect of thermal annealing on time-evolution of a thin film morphology. We conclude by formulating natural extensions of our framework to characterize crystallinity and anisotropy of the morphology using the framework.

\end{abstract}


\end{frontmatter}


\section{Introduction}
\label{ch:intro}
Organic solar cells (OSC) fabricated from polymer-fullerene blends~\cite{ATT09,CPH04,DSB09,HS04,KF08,TF07} represent a promising low-cost, rapidly deployable strategy for harnessing solar energy~\cite{PRB09}. The possibility of large-area fabrication on flexible substrates makes these devices highly attractive for ubiquitous solar-electric conversion. The past decade has witnessed considerable advances in OSC technology, both from the perspective of understanding the underlying physical processes and concurrent improvement in efficiencies (from PCE below 3\%~\cite{SBS01} to currently the highest reported efficiency 8.13\%~\cite{GreenWarta2011} obtained at the laboratory scale). 

Along with synthesis of new organic semiconductors\footnote{e.g., PBDTTT~\cite{ChenLi2009}, PCDTBT~\cite{PRB09}, PTB systems~\cite{LFW09}, ICBA~\cite{HCH10}.} and development of new device architectures\footnote{e.g., inverted architecture~\cite{HYZ09,WOS06}, tandem cells~\cite{CRH10,KLC07}.}, improvements in efficiencies have come from optimizing material processing to improve morphology. Improving morphology is crucial because all physical processes within an OSC (and thus the device performance) are a strong-function of thin film morphology. Common processing techniques such as thermal annealing, solvent annealing, and utilization of solvent additives affect the morphology~\cite{LSH05,MYG05,PeetBazan2007}.
Morphology is also affected by fabrication conditions like solvent evaporation rate, substrate surface, solvent type, and donor-acceptor blend ratio, among others.

\begin{figure}[h]
\parbox{0.49\textwidth}{(a)}
\parbox{0.49\textwidth}{(b)}\\
\parbox{0.49\textwidth}{
\includegraphics[width=0.48\textwidth]{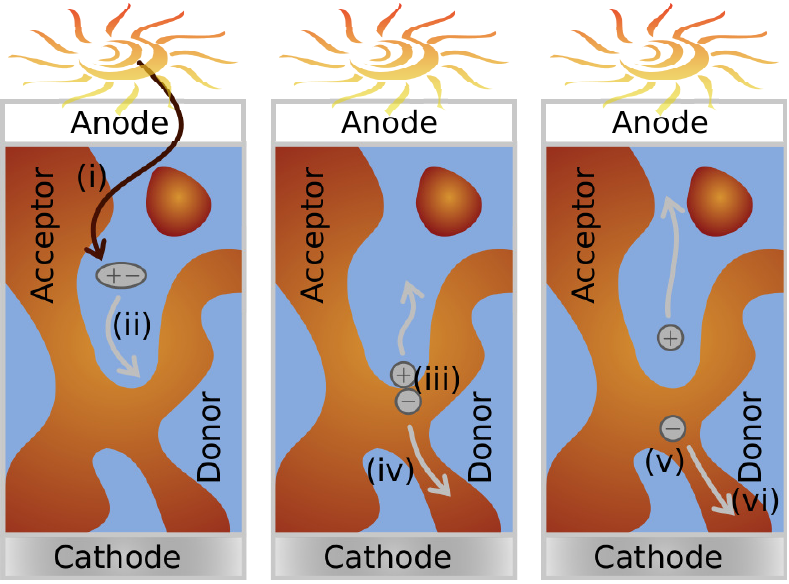} 
}
\parbox{0.49\textwidth}{
\includegraphics[width=0.49\textwidth]{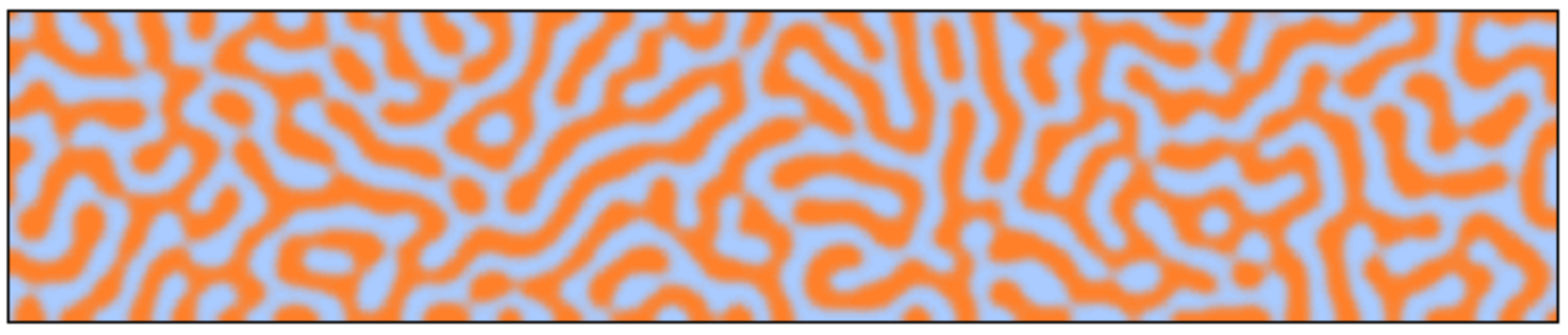}\\ 
\includegraphics[width=0.49\textwidth]{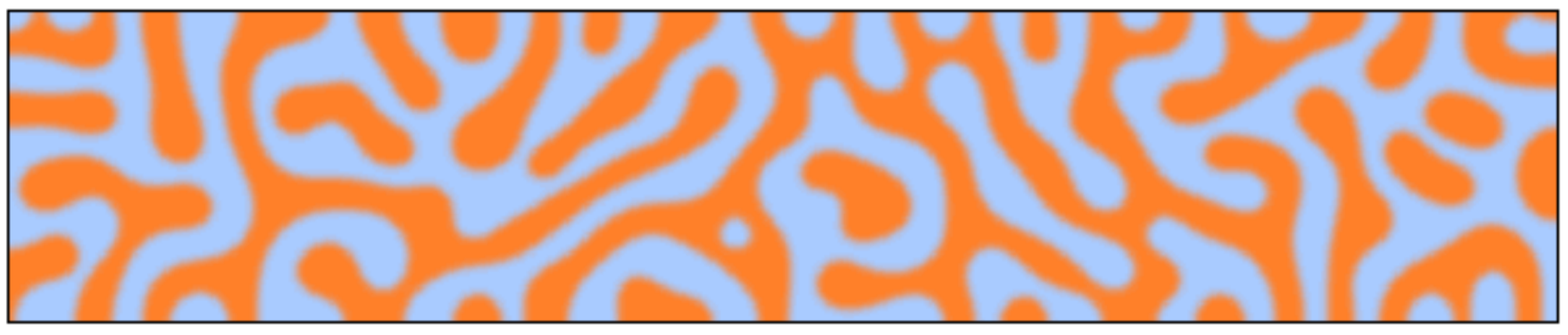}
} 
\caption{(view in color) Link between bulk heterojunction morphology and photocurrent generation. Figure on left~(a) lists the physical processes affected by the morphology: (i) exciton generation,
(ii) exciton diffusion, (iii) exciton dissociation, (iv) charges separation,
(v) charge transport, (vi) extraction of charges. Figure on right~(b) shows two similar looking morphologies with very different features -- number of isolated islands and regions connected to electrodes, tortuosity -- which will lead to different device characteristics.}
\label{fig:photovoltaic}
\end{figure}

{\bf The photovoltaic behavior and its dependence on (bulk heterojunction) morphology}: 
Most OSCs have bulk-heterojunction architecture (BHJ)~\cite{PRB09,LSH05,PeetBazan2007}. BHJ consists of an electron donating polymer and electron accepting fullerene-derivatives usually intermixed with a phase separation scale of the order of exciton diffusion length.
The final morphology of electron-donor and electron-acceptor regions in the device strongly affects the power conversion efficiency of OSCs. This has its origin in the operating principle of organic solar cells (see Figure~\ref{fig:photovoltaic}), which can be broadly classified into three stages: (1) light absorption and exciton generation, (2) exciton diffusion to the donor-acceptor interface and dissociation into free charges, and (3) charge transport towards the electrodes.

Light absorption by photoactive material leads to creation of excitons. These excitons have a finite lifetime (before recombining due to Coulomb attraction) during which they can diffuse. Excitons dissociate to form useful free charges \emph{only at the donor-acceptor interface}~\cite{HB08}. Thus, only those excitons that are created a short distance from the interface are able to reach the donor-acceptor interface and dissociate into free charge carriers. The final step of photocurrent generation is transport of free charge carriers to respective electrodes. 
Electrons travel only through electron-acceptor rich regions to the cathode, while holes travel only through electron-donor rich regions to the anode. Consequently, if charges are created in islands or cul-de-sacs with no direct connection to the appropriate electrode, they eventually recombine. Therefore  continuous pathways of electron-donor and electron-acceptor regions directly connected to appropriate electrodes are necessary for charge transport and extraction. Thus, every stage of the photovoltaic process is extensively affected by the morphology of the thin film.

{\bf Necessity for morphology descriptors:} There is significant merit to develop comprehensive morphology descriptors to understand and quantify the role that morphology plays in performance. This notion is primarily motivated by the following reasons:
\begin{itemize}
	\item A major challenge for fabricating high efficiency OSCs by tailoring the morphology is {\it rather weak control over fabrication process}. This is due to the large number of process variables (like evaporation rate, substrate effect, temperature, and solvent environment) and system variables (like blend ratio, solvent type, and solvent blends), which affect the morphological evolution. A detailed knowledge of process--structure--property relationships could become the cornerstone for optimizing fabrication conditions and realizing high-efficiency OSCs. To link structure and property, it is crucial to extract quantitative and physically meaningful morphology descriptors from the morphology.
	\item Progress in experimental characterization~\cite{AHM09,BSWL09,MLC09} and computational modeling~\cite{TsigeGrest2005,PuemansForrest2003,WG10a} now allow in-situ and in-silico 3D reconstruction and imaging of the thin film morphology with unprecedented accuracy and resolution. A computational framework capable of constructing morphology descriptors efficiently, can enable rapid and effective assimilation of large morphology (both temporal and spatial) data sets.
\end{itemize}
 
{\bf State-of-the-art morphology descriptors for OSC:} Nanomorphology descriptors are almost absent in the context of OSCs. There have only been a few attempts to characterize selected morphological properties affecting the final efficiency using (i) the autocorrelation function~\cite{MYH07,MLC09}; (ii) the interfacial area between electron-donor material and electron-acceptor material~\cite{LYK08,MSL10}, and (iii) the fraction of volume connected to the relevant electrode~\cite{BSW09}. While these quantities give important insight into the effect of morphology on device physics, they essentially provide only coarse information. 

The averaged feature size of morphology can be extracted from the autocorrelation function. The feature size allows to assess the inclination of excitons to reach the interface.\footnote{In practice the feature size is linked with the ``exciton-diffusion length'' which is around $10nm$ for P3HT~\cite{Shaw2008} (the best studied conjugated polymer in the context of OSCs). When the averaged feature size is comparable or smaller than exciton diffusion length, there is a high probability for high exciton dissociation. Otherwise, only a small fraction of excitons can physically reach the interface and contribute to current generation.} Although insightful, this descriptor provides only coarse, averaged information. In particular, the averaged domain size does not distinguish between electron-donor and electron-acceptor regions (\emph{lack of selective quantification}). 

The fraction of volume connected to relevant electrodes provides important insight into the charge transport by quantifying the useful regions of domain. 
However, this descriptor does not provide any quantitative information on the length or tortuosity of paths in the morphology. Quantifying the tortuosity of the morphology (of both electron-donor and electron-acceptor regions) is of particular interest, due to the finite life time of charges and the differences in the charge mobilities of electrons and holes.

{\bf Graph-based approach:}  In this paper, we provide a comprehensive suite of physically meaningful morphology descriptors, which reflect the complex nature of the BHJ and the underlying device physics. We present a novel graph-based framework to efficiently construct this comprehensive suite. This approach is motivated by the equivalence between a discretized 2D/3D morphology image and a {\it labeled, weighted, undirected graph}. These morphology descriptors are further classified according to the physical subprocesses of the photovoltaic process: exciton diffusion, exciton dissociation into free charge carriers, and charge transport. We utilize this approach to pose six device-performance relevant questions to exhaustively characterize the morphology. We subsequently construct estimates and rigorous upper bounds of various efficiencies. The approach is showcased by characterizing the effect of thermal annealing on time-evolution of a thin film morphology. The developed framework of morphology descriptors is generic and can be applied in other areas, e.g., percolation pathways in geomechanics~\cite{Shaw2005}, porous media~\cite{Torquato2002,Hunt2009}, and drug release from polymeric membranes~\cite{SKP07}.

This paper is organized as follows. In Section 2, a basic taxonomy of morphology descriptors is introduced and a list of questions that augment morphology characterization is formulated. Sections 3 and 4 motivate the graph-based approach. This is achieved by detailing the equivalence between a digitalized morphology and a labeled, weighted, undirected graph. In Section 5, we formulate a set of questions efficiently answered using graph-based algorithms. We subsequently utilize these questions to estimate the upper bounds on constitutive efficiencies in Section 6. Results and conclusions are presented in the last two sections.

\section{A taxonomy of physically meaningful morphology descriptors}
\label{ch:Mo}

We broadly define four groups of morphology descriptors to characterize the nano-morphology: descriptors quantifying light absorption, exciton diffusion, exciton dissociation and charge transport.   
\begin{itemize}
\item \emph{Light absorption:} In most BHJs, light is absorbed by the electron-donor material to create excitons.\footnote{Electron acceptors also absorb light to create excitons. However, in most cases light absorption and exciton harvesting are very weak compared to donor material. This is the case for PC$_{61}$BM -- the most common acceptor~\cite{BurkhardMcGehee2009}.} 
Thus, a natural descriptor is  the fraction of electron-donor material within the active layer of the device. 
\item \emph{Exciton diffusion:} The key metric that affects exciton diffusion is simply the distance to the nearest donor-acceptor interface. An ideal morphology would be a structure where the entire photoactive material is distributed within a distance shorter than the exciton diffusion length from the donor-acceptor interface. 
\item \emph{Exciton dissociation:} To increase the number of the separated charges, the length of the interface should be maximized. Thus, the most natural morphology descriptor reflecting the nature of exciton dissociation is the interfacial area.
\item \emph{Charge transport:} An ideal morphology should have following characteristics for good charge transport:
	\begin{itemize}
		\item[(i)] \emph{A percolating network} of electron-donor and electron-acceptor materials. The morphology should not have any ``islands'' or ``cul-de-sacs'' to minimize recombination. In other words, all donor and acceptor domains should be connected to relevant electrodes. 
		\item[(ii)] \emph{Short paths} to both electrodes: This is important to minimize the possibility of recombination and also accounts for mean free path  of carriers.
		\item[(iii)] \emph{Balanced paths} to avoid charge accumulation, that is, the electron and hole path lengths should be similar. Furthermore, the paths must be weighted by the charge mobilities to account for different transport velocities of electrons and holes.
	\end{itemize}
	The above characterizations of an ideal morphology motivates the following descriptors: the fraction of useful domains, the distance charges travel from interface to electrodes via electron-donor or electron-acceptor regions, and the fraction of interface with complementary paths to both electrodes.
\end{itemize}

Such a categorizations provide a clear physics-based rationale and provides succinct guidance for establishing links between morphology and each constitutive subprocess in the photovoltaic process within an OSC. In the sequel, we develop efficient graph-based techniques to construct a suite of descriptors listed above. We formulate six questions related to the aforementioned  subprocesses. These questions provide insight into the factors affecting the final efficiency of the device:
\begin{enumerate}
\item[Q1] What is the fraction of light absorbing material? (affects absorption efficiency -- $\eta_{abs}$)
\item[Q2] What is the fraction of electron-donor material whose distance to the donor-acceptor interface is within a given range ? (affects exciton dissociation efficiency -- $\eta_{diss}$)
\item[Q3] What is the interfacial area between donor and acceptor
regions? (affects exciton dissociation efficiency -- $\eta_{diss}$)
\item[Q4a] What is the fraction of donor domain connected to anode? (affects hole transport and extraction -- $\eta_{out}$)
\item[Q4b] What is the fraction of acceptor domain connected to cathode? (affects electron transport and extraction -- $\eta_{out}$)
\item[Q5a] What is the distance from donor-acceptor interface to the anode via donor domain only? (affects hole transport and extraction -- $\eta_{out}$)
\item[Q5b] What is the distance from donor-acceptor interface to the cathode via acceptor domain only? (affects electron transport and extraction -- $\eta_{out}$)
\item[Q6] What is the interface fraction which has complementary paths to both electrodes? (affects charges transport and extraction -- $\eta_{out}$)
\end{enumerate}


\section{Digitized morphology as labeled, weighted, undirected graphs}
\label{sec:DMLWUG}

\begin{figure}[h]
\parbox{0.24\textwidth}{(a)}
\parbox{0.24\textwidth}{(b)}
\parbox{0.24\textwidth}{(c)}
\parbox{0.24\textwidth}{(d)}\\
\includegraphics[width=1\textwidth]{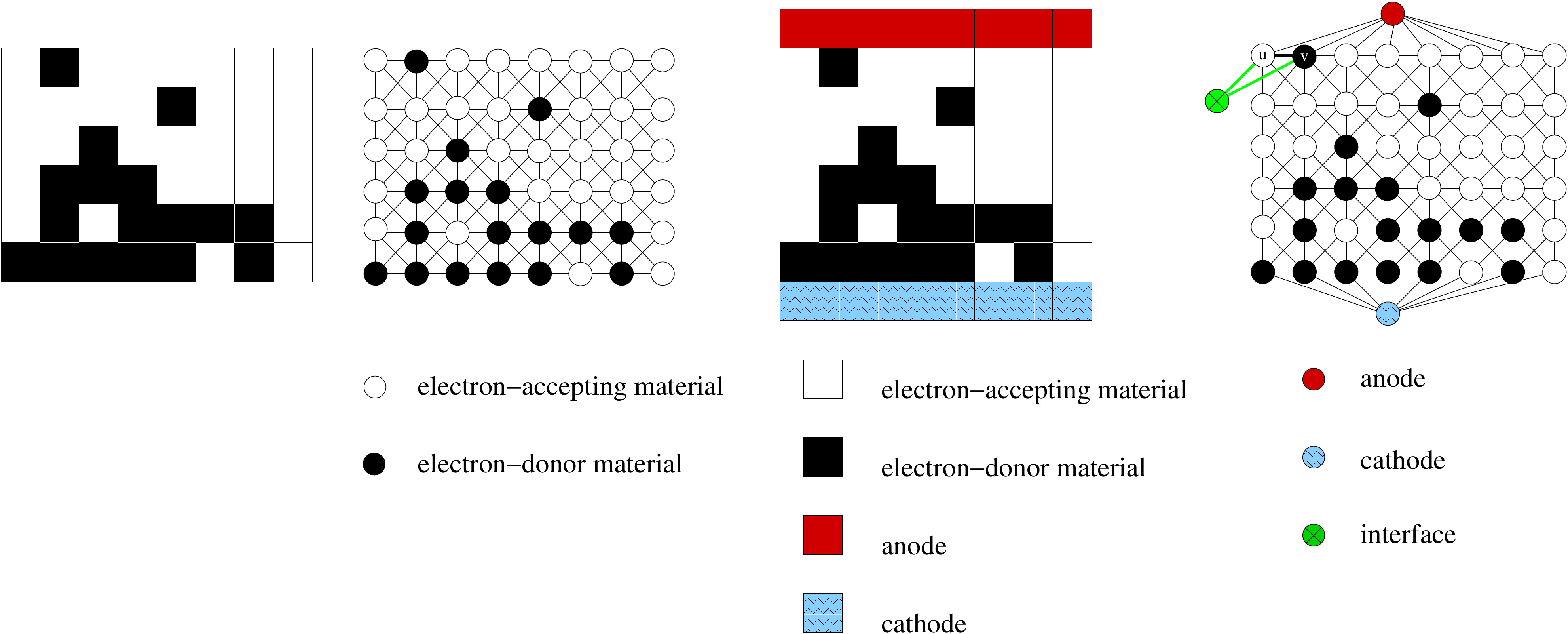}
\caption{Graph construction for two-phase morphology. (a) A simple digitized morphology representing the BHJ. (b) Each unit element (pixel) of BHJ is represented as a vertex in the graph with corresponding color assigned. Local neighborhood is used to connect vertices through edges. (c) BHJ sandwiched between two electrodes. (d) Three meta vertices are added to the graph and represent anode, cathode, and interface. Anode vertex is connected to all top vertices via additional edges. Cathode vertex is connected to all bottom vertices. We represent interface as one meta vertex. Every vertex pair across the interface is connected to this meta vertex. An example insertion of such edges is shown in~(d). Note that not all edges connected to the interface are shown in the figure.}
\label{MS2GP}
\end{figure}

The following two key observations motivate the graph-based approach:
\begin{enumerate}
\item  The information in a digitized morphology can be represented
using an appropriately defined weighted, undirected graph. Each pixel (or voxel) becomes a graph vertex with a label denoting its phase. Each vertex is connected to its neighboring vertices through edges whose weight depend on the distance between the vertices. This is illustrated using a simple morphology in Figure~\ref{MS2GP}. A detailed discussion is provided in Section~\ref{subsec:GraphConst}.
\item Many questions on morphology characterization can be recast as
questions on the structural properties of an appropriately defined
subgraph of the above graph. In particular, the following fundamental
graph questions occur repeatedly in answering questions Q1-Q6.
	\begin{itemize} 
	\item {\it Identifying distinct components in a graph:} These
correspond to the number of distinct donor or acceptor domains, connectivity of these domains to the electrodes.
	\item {\it Computing shortest distances between a pair of vertices in a graph.}
	\end{itemize}
Both of the above questions on graphs have been well studied and near-optimal
algorithms for them are available~\cite{CormenBook}. The graph-based
approach lets us leverage this prior work to solve the morphology
characterization problem.
\end{enumerate}

These observations in conjunction with the descriptors defined in Section~\ref{ch:Mo} result in the formulation of highly efficient algorithms to characterize the BHJ morphology.

\section{Using Graph Theory to characterize BHJ morphology}

In this section, we introduce some basic definitions about graphs and
formalize the equivalence between a digitized morphology and a graph.

\subsection{Basic definitions}

\begin{itemize}
\item An undirected graph $G=(V,E)$ is defined by a set of vertices,
$V$, and a set of edges, $E$, where each edge in $E$ is an unordered
pair of vertices drawn from $V$.

\item A weighted undirected graph $G=(V,E,W)$ is an undirected graph
$(V,E)$ with an associated weight function, $W: E \rightarrow {\mathbb R}_{+}$, that assigns a non-negative 
real weight to each edge in $E$.

\item A labeled weighted undirected graph $G=(V,E,W, L)$ is a weighted
undirected graph $(V,E,W)$ with an associated labeling function, $L$,
that assigns a label to each vertex in $V$. In this work, we label each vertex by a color.

\item A path between a source vertex, $s \in V$, and a target vertex,
$t \in V$ is a sequence $p=[v_0, v_1, \dots v_i \dots v_k]$ of
vertices such that $v_o=s$, $v_k=t$ and 
for each $i$ from $0$ to $i-1$, vertices $v_i$ and $v_{i+1}$ are
adjacent in $G$. 
The length of path $p$ is $\sum_{i=0}^{k-1}w(e(v_i,v_{i+1}))$.

\item A shortest path between a source vertex $s \in V$ and a
target vertex $t \in V$ is a path between $s$ and $t$ that is of the
shortest length among all paths between $s$ and $t$ in $G$. The
distance between vertices $s$ and $t$ in $G$ is the length of a shortest
path between $s$ and $t$ in $G$. If no such path exists, the distance is defined as infinity.  
Note that the shortest path between a pair of vertices need not be unique, but the distance between them
is unique.

\item A subgraph of $G$ is a graph $G' = (V', E')$ such that 
$V' \subseteq V$ and $E' \subseteq E$. A vertex-induced subgraph 
on vertex set $V' \subseteq V$ is the maximal subgraph with the
vertex set $V'$.

\item A graph $G$ is connected if there is a path between any pair of
vertices in $G$. A connected component $C$ in $G$ is a maximal
connected subgraph of $G$. 
\end{itemize}

\subsection{Graph construction}
\label{subsec:GraphConst}

The graph construction starts from a given digitized (two-phase)
morphology. Such a morphology can be obtained either from
experiments~\cite{MLC09,BSWL09} or from numerical
simulations~\cite{TsigeGrest2005,PuemansForrest2003,WG10a}. 

Given a two-phase morphology, we construct a labeled weighted
undirected graph $G = (V,E,W,L)$, as follows.

There is one vertex $v \in V$ corresponding to each pixel in the morphology
(voxel in 3D reconstruction). 
For each vertex, we use the inherent structure of the morphology
(i.e., pixels are located on a uniform lattice) to define the edge set
$E$. There is an edge between each pair of vertices that correspond to
neighboring pixels (voxels) in the morphology. For example, in two
dimensions, a pixel can have 8 neighboring pixels, hence a vertex
corresponding to a pixel can have up to 8 neighbors in the graph.
Each edge $e=(u,v) \in E$ is assigned a weight $W(e)$ equal to the Euclidean
distance between the pixels corresponding to $u$ and $v$ in the
morphology. First order neighbors one lattice distance away have
an edge weight of $1$, and second order neighbors  $\sqrt{2}$
lattice units away have an edge $\sqrt{2}$ weight. \footnote{For the 3D
case, third order neighbors must be considered.}

Each vertex $v \in V$ is assigned a label $L(v)$ which is ``black'' or
``white'', depending on the phase of the pixel corresponding to
$v$. Black represents donor-material, white represents
acceptor-material. More generally, a label may represent phase,
crystallinity or any other morphological characteristic.

We add three more vertices to the graph, corresponding to the anode,
the cathode, and the interface, respectively (see
Figure~\ref{MS2GP} (c) and (d)). There is an edge of weight $1$ between the anode
(cathode) and each vertex $v \in V$ that corresponds to a pixel which
is physically adjacent to the anode (cathode). There is an edge of
weight $0.5$ from the interface vertex to each black (white) vertex in
$v \in V$ such that $v$ has an edge of weight $1$ to a white (black)
vertex. The anode, cathode, and interface vertices have labels
``anode'', ``cathode'', and ``interface'', respectively.
Adding these vertices to the graph allows for easy morphology
interrogation (e.g., estimation of graph distances from any point of
the domain to the electrodes).

{\bf Remark 1}: The edge weights provide a measure of distance between
connected vertices. We use nondimensional units to represent these
edge weights (or distances). This can be easily linked with physical
length using a scale (converting length to pixels/voxels).

This conversion of a discretized morphology into a graph has several
advantages:
\begin{enumerate}
\item A graph is independent of the dimensionality of the original morphology considered (2D and 3D). Thus, techniques developed on graphs can be applied irrespective of the dimensionality. 
\item Most graph algorithms are highly optimized and their computational complexity well studied, thus ensuring fast and efficient morphology characterization. Moreover, they are scalable, allowing the study of large data sets.
\item Periodicity on boundaries can be trivially introduced (by appropriate edges construction).
\item Distances in the graph can be easily converted to physical units.
\item A graph-based approach is generic and can be naturally extended. In particular, crystallinity and anisotropy effects can be investigated by introducing additional labels.
\end{enumerate}

\subsection{Two basic operations on graphs}
\label{subsec:connComp}

As alluded to in Section~\ref{sec:DMLWUG}, morphology characterization
can be posed on the basis of two general procedures on graphs: (1)~finding
connected components, and (2)~computing the shortest paths in the graph. We
describe them below and use them to construct algorithms in
Section~\ref{sec:char:BHJ}.

\subsubsection{Identifying connected components of a graph}

Efficient computation of the connected components of a graph is a
powerful tool, when used with proper filtering of the graph. Graph
filtering involves virtual masking of edges to retain only those edges
satisfying specific properties.

\begin{figure}
\begin{center}\includegraphics[width=0.3\textwidth]{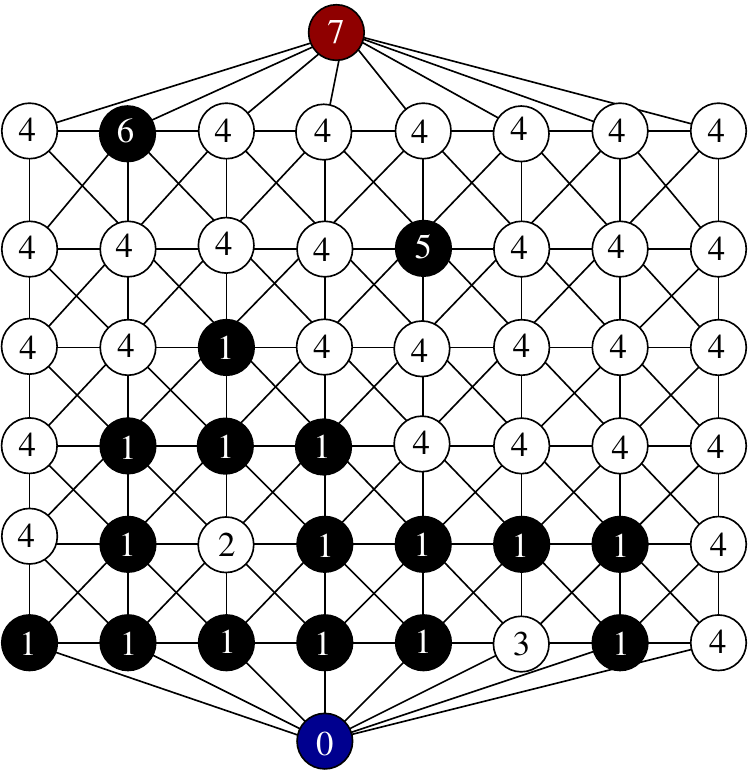}\end{center}
\caption{Example of a simple graph with indices of components marked (there are 8 connected components numbered $0\dots7$). Finding the connected components provides the following information: (i) number of black components: 3, (ii) number of white components: 3, (iii) number of black components connected to top: 1 (iv) number of white components connected to bottom: 2. (v) number of black vertices connected to top: 1, (vi) number of white vertices connected to bottom: 30, (vii) number of black vertices isolated from top: 16, (viii) number of white vertices isolated from bottom: 1.}
\label{fig:conn:components}
\end{figure}

For example, in organic BHJ, excitons and charges can move within
domains of one kind (donors or acceptors). Therefore, it is of
interest to identify these subdomains (of donors or acceptors) in the
morphology.  This can be accomplished in two stages: graph filtering
and finding connected components in the filtered graph. The filtering
step consists of virtually masking edges between vertices of different
colors.  As the outcome of this algorithm, each vertex of the graph
has an assigned index of the corresponding component. Figure~\ref{fig:conn:components} illustrates a simple example
with the connected components marked. Each vertex has an assigned
index representing the component it belongs.

\subsubsection{Tortuosity of morphology: shortest path in the graph}


Efficient computation of the shortest path in a graph is useful to estimate tortuosity of paths in a BHJ morphology. In particular, this can be used to estimate the path lengths of (a) excitons as they diffuse towards donor-acceptor interface, and (b) charges as they move via tortuous domains to a specific electrode. 
The single source shortest path problem can be efficiently solved using the Dijkstra algorithm~\cite{Dijkstra1959,CormenBook}.

\section{Six questions to characterize bulk heterojunction morphology}
\label{sec:char:BHJ}
We formulate a set of questions which provide a hierarchical characterization of the BHJ morphology. We utilize graph theory concepts (particularly, connected components and graph distances) to construct efficient algorithms to answer these questions. 

\subsection {Q1: Estimate the fraction of light absorbing material}

In a BHJ, light is usually absorbed by the electron-donor material. A simple estimate of the fraction of electron-donor material provides some insight into the amount of incident light absorbed. We include this trivial estimate to demonstrate that different characterizations can be performed easily using the graph approach. 
	
\noindent\framebox[\textwidth]{
	\parbox{0.85\textwidth}{
		\begin{enumerate}
	  	\item[Input:] Given labeled, weighted, undirected graph, $G=(V,E,W,L)$.
	  	\item Identify set $B$ of all (electron donor) black vertices.
	  	\item[Output:] $|B|/|V|$, where $|.|$ is the cardinality of the set.
	  \end{enumerate}
	 }
}


\subsection {Q2: Fraction of photoactive material whose distance to the donor-acceptor interface is a within given range (particularly within the exciton diffusion length)}

Light is absorbed to generate excitons. These excitons diffuse to the interface, only where they dissociate into charges. Excitons have a limited lifetime or a maximum distance called exciton diffusion length, after which they recombine. Therefore, we want to find the fraction of photoactive material whose distance to the interface is within this given range.
This descriptor is constructed in three steps. First, we construct the subgraph induced by a set of black and interface vertices. Next, using Dijkstra's algorithm the shortest paths between interface vertex and all other vertices in the induced subgraph are determined. The identification of all vertices whose path lengths are shorter than the given distance is the final stage of this procedure. Formally:

\noindent\framebox[\textwidth]{
	\parbox{0.85\textwidth}{
		\begin{enumerate}
		  \item[Input:] Given labeled, weighted, undirected graph $G=(V,E,W,L)$.
		  \item Construct vertex-induced subgraph, $G'=(V',E')$, where: $V'$ is a set of all black and green vertices in $V$, $E'$ is a set of all edges between vertices in $V'$.
		  \item Find all shortest paths in $G'$ from interface vertex $u$ $(L(u) = green)$. 
		  \item Identify vertex set $V_d=\{v \in V'\;|\;$ (shortest\_path($v$) $<d$) $\land (L(v) = black)\}$. Shortest\_path ($v$) denotes the distance between $u$ and $v$ in $G'$.
		  \item[Output:] $|V_d|/|B|$, where $|B|$ is the cardinality of a set that consists of black vertices.  
		\end{enumerate}
	}
}
By varying the distance, $d$, a histogram (or the probability mass function) of the shortest distances from donor to the interface can be computed. This is particularly important for a broader analysis of BHJ morphology; for example, when exciton diffusion length is not known a priori or if it depends on operational conditions, e.g., temperature. 

\begin{figure}
\begin{center}
\parbox{0.25\textwidth}{\centering
\includegraphics[width=0.15\textwidth]{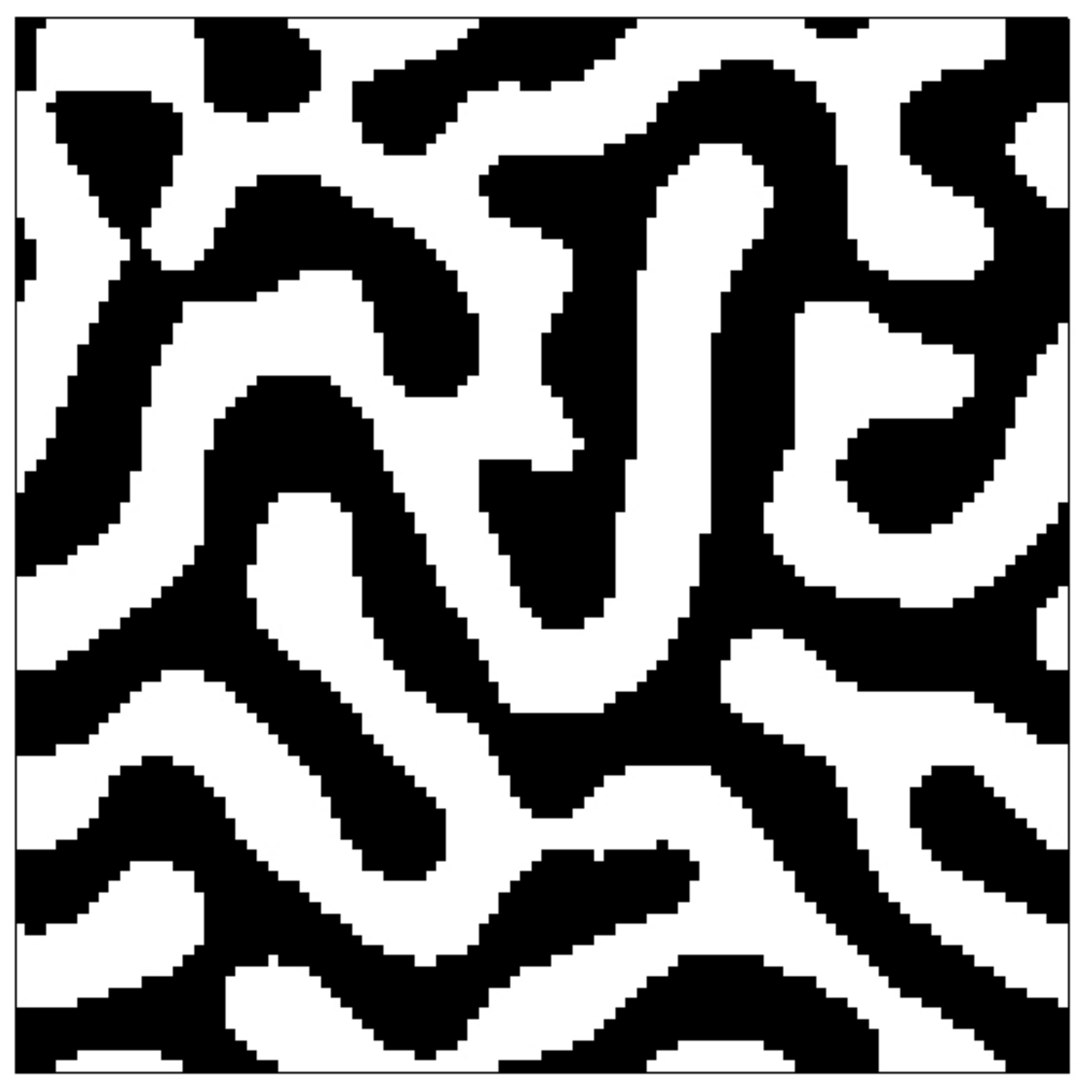} 
}
\parbox{0.25\textwidth}{\centering
\includegraphics[width=0.15\textwidth]{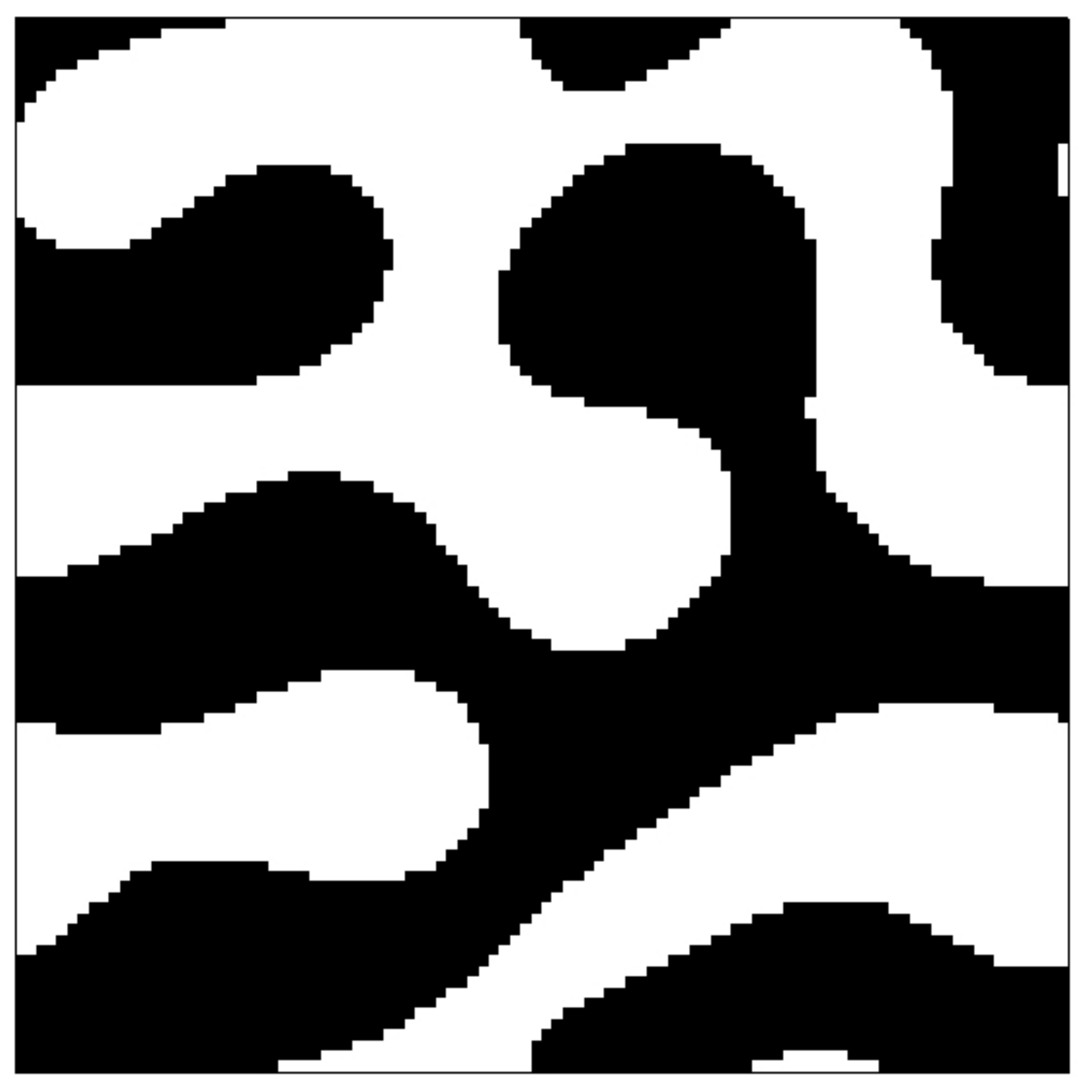}
}
\parbox{0.25\textwidth}{\centering
\includegraphics[width=0.15\textwidth]{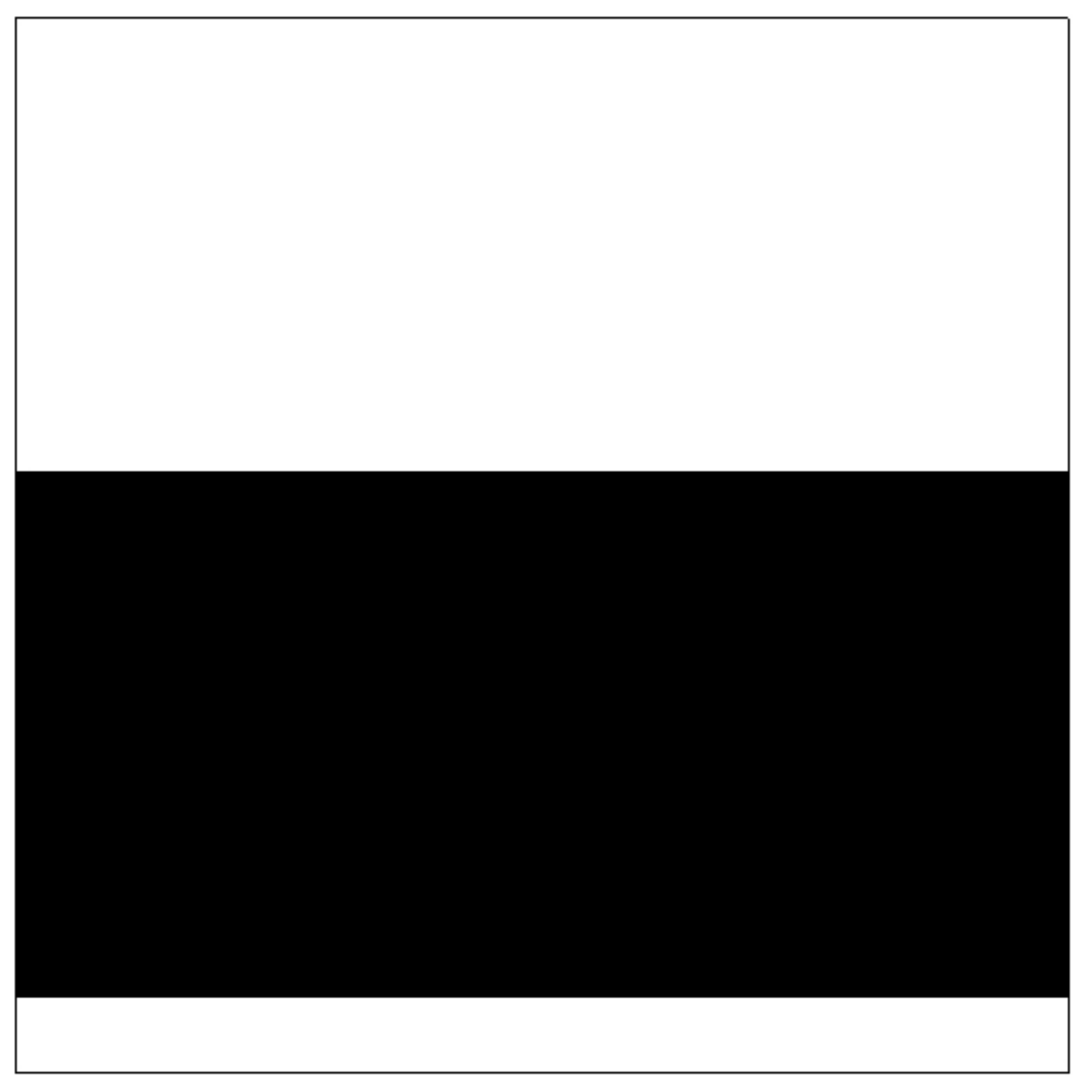}
}\\
\includegraphics[width=0.25\textwidth]{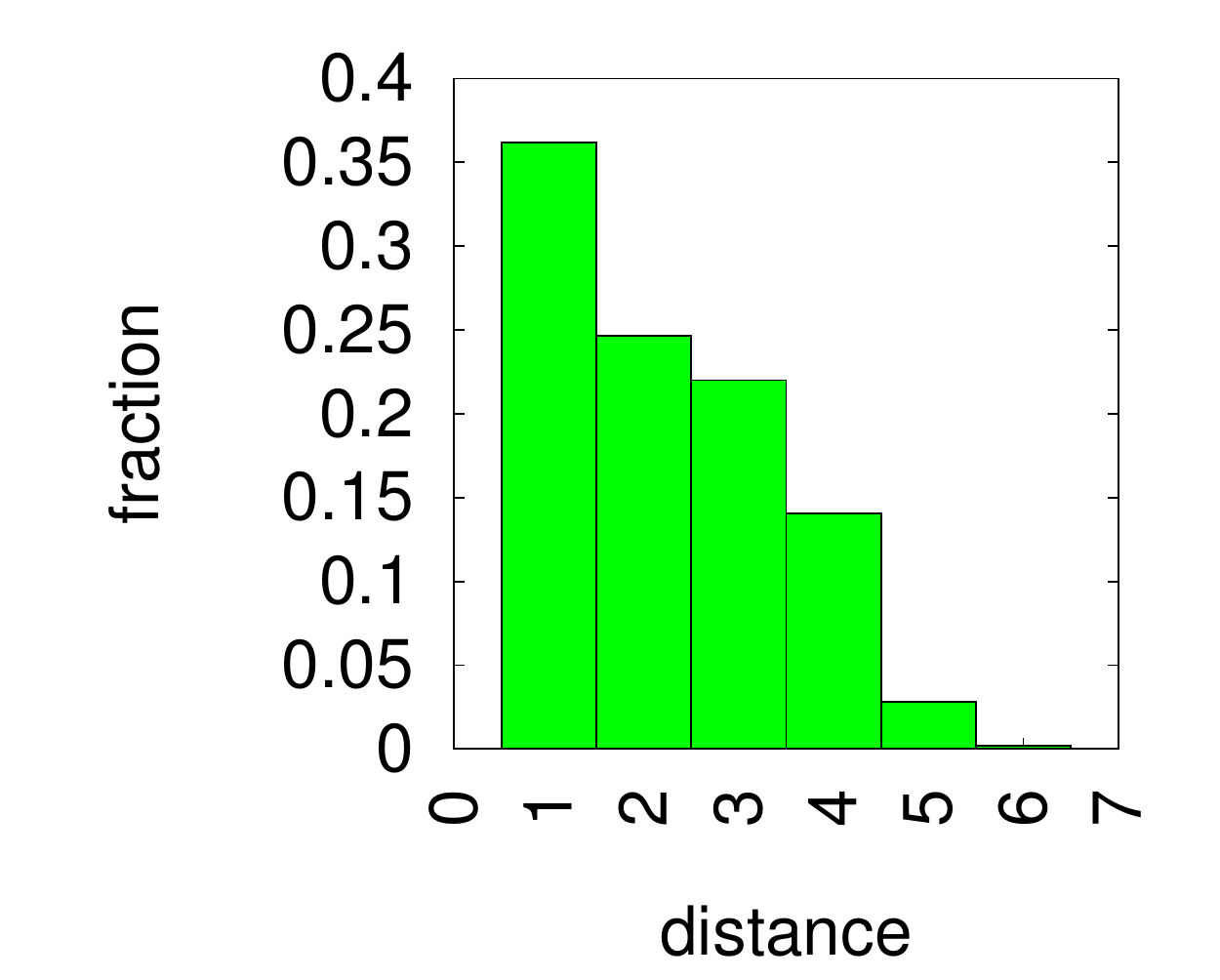} 
\includegraphics[width=0.25\textwidth]{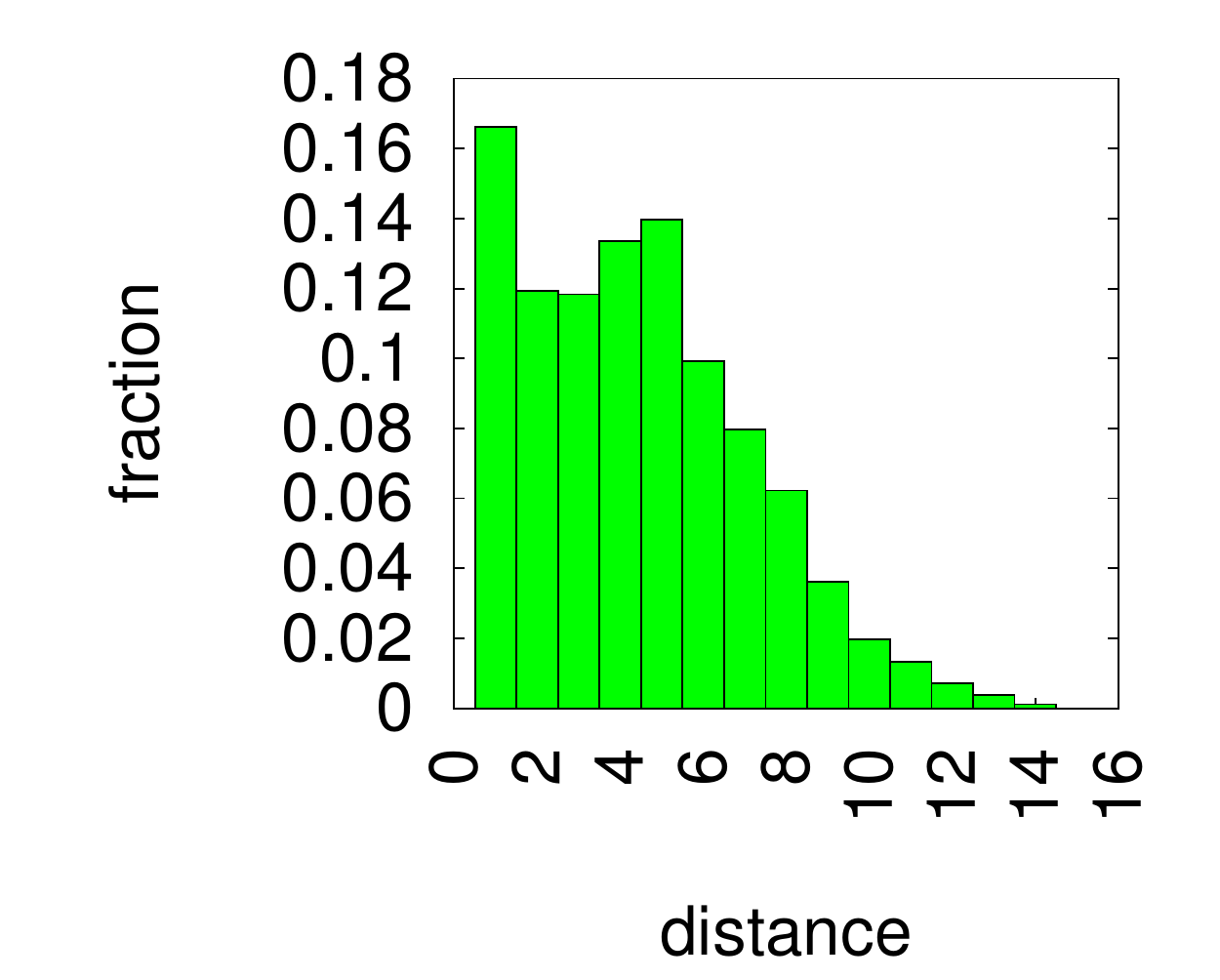} 
\includegraphics[width=0.25\textwidth]{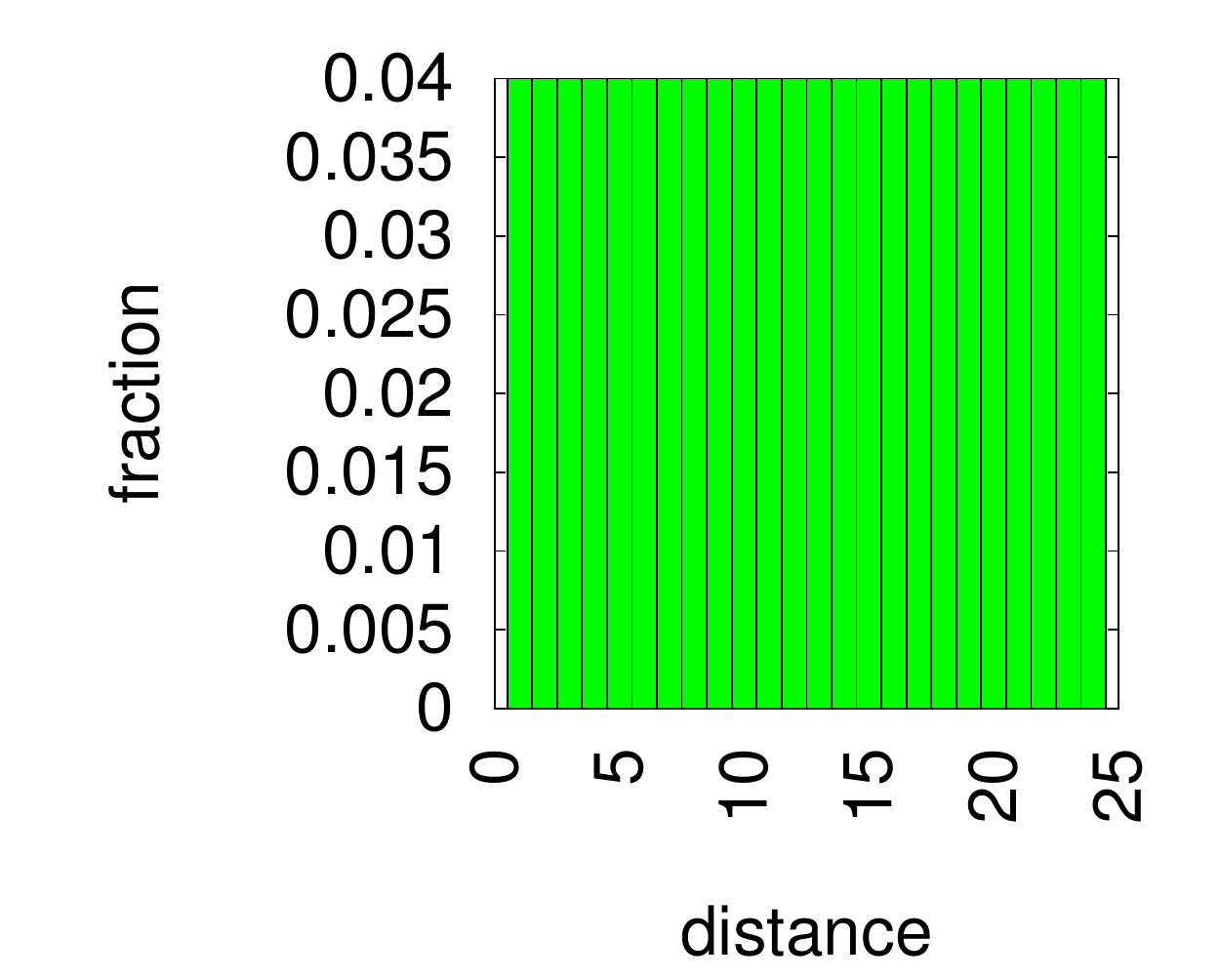} 
\\
\includegraphics[width=0.25\textwidth]{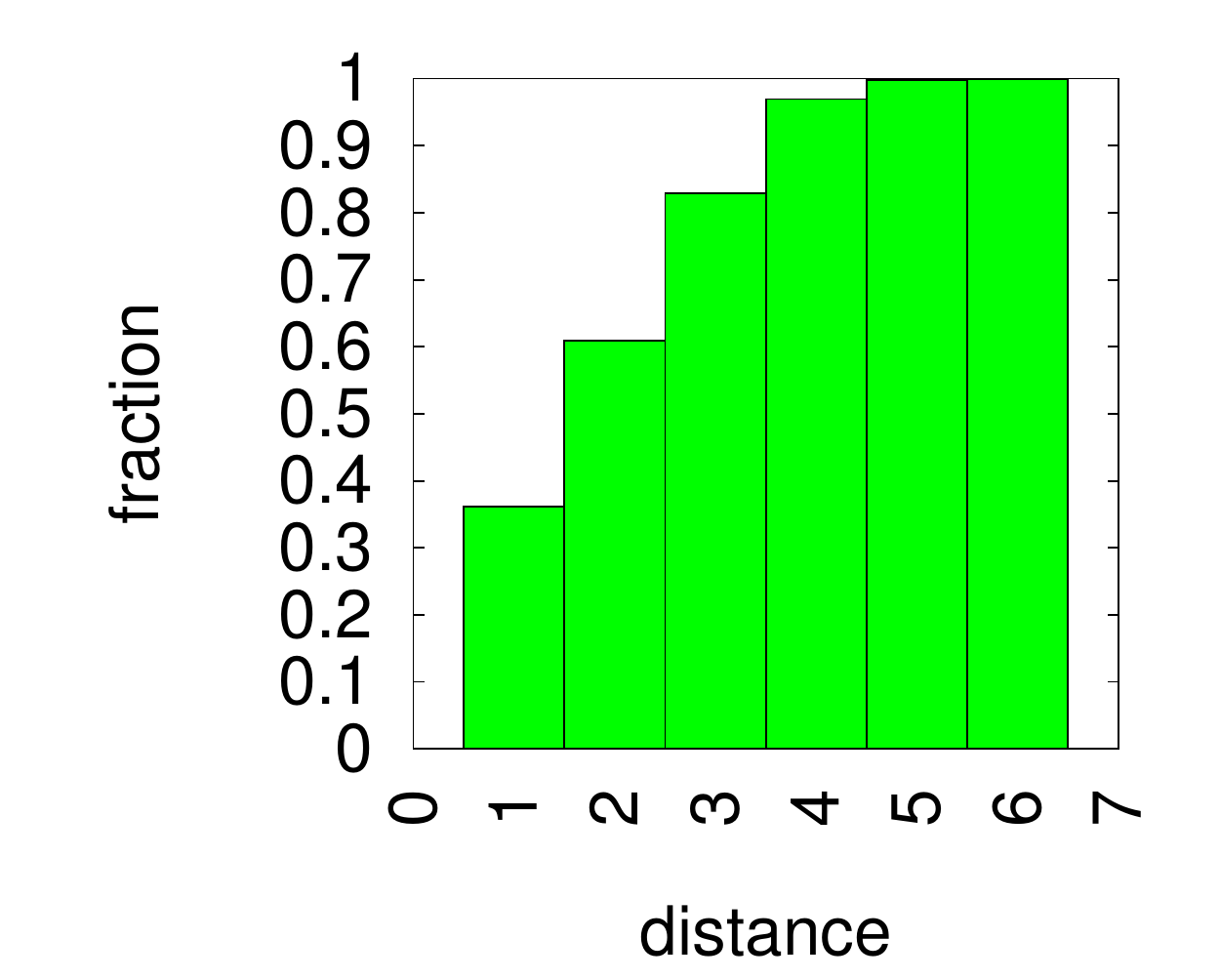} 
\includegraphics[width=0.25\textwidth]{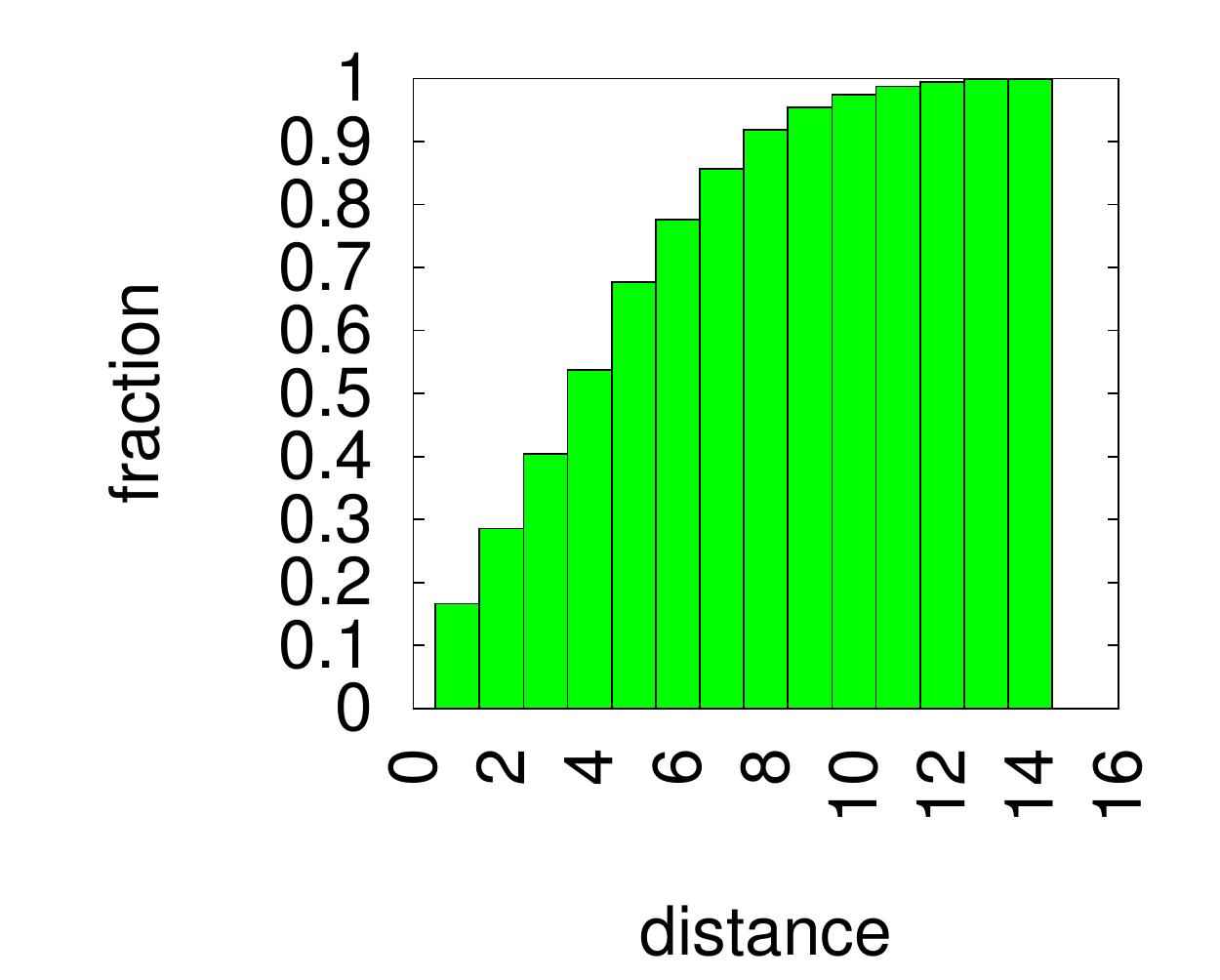} 
\includegraphics[width=0.25\textwidth]{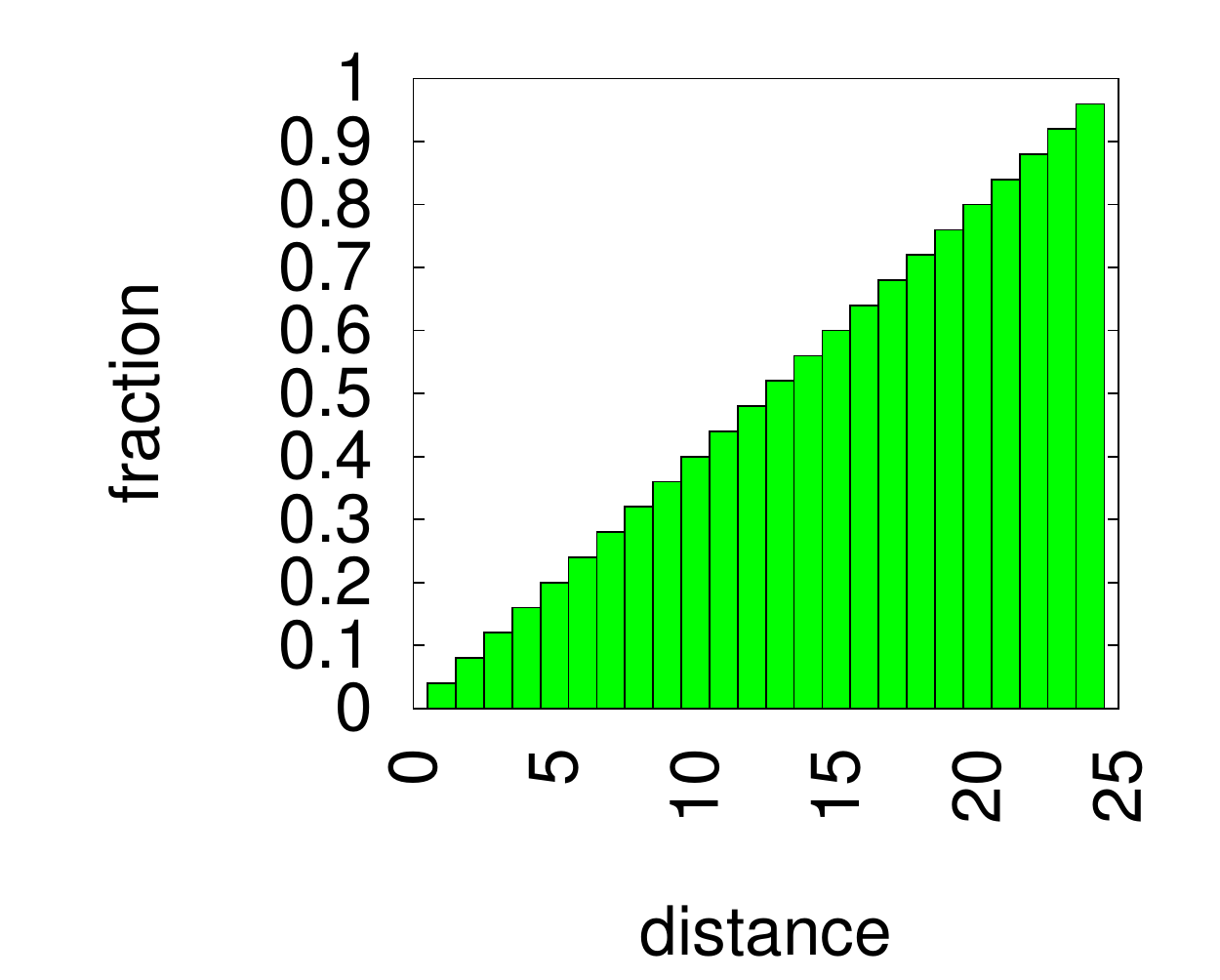} 
\\
\includegraphics[width=0.28\textwidth]{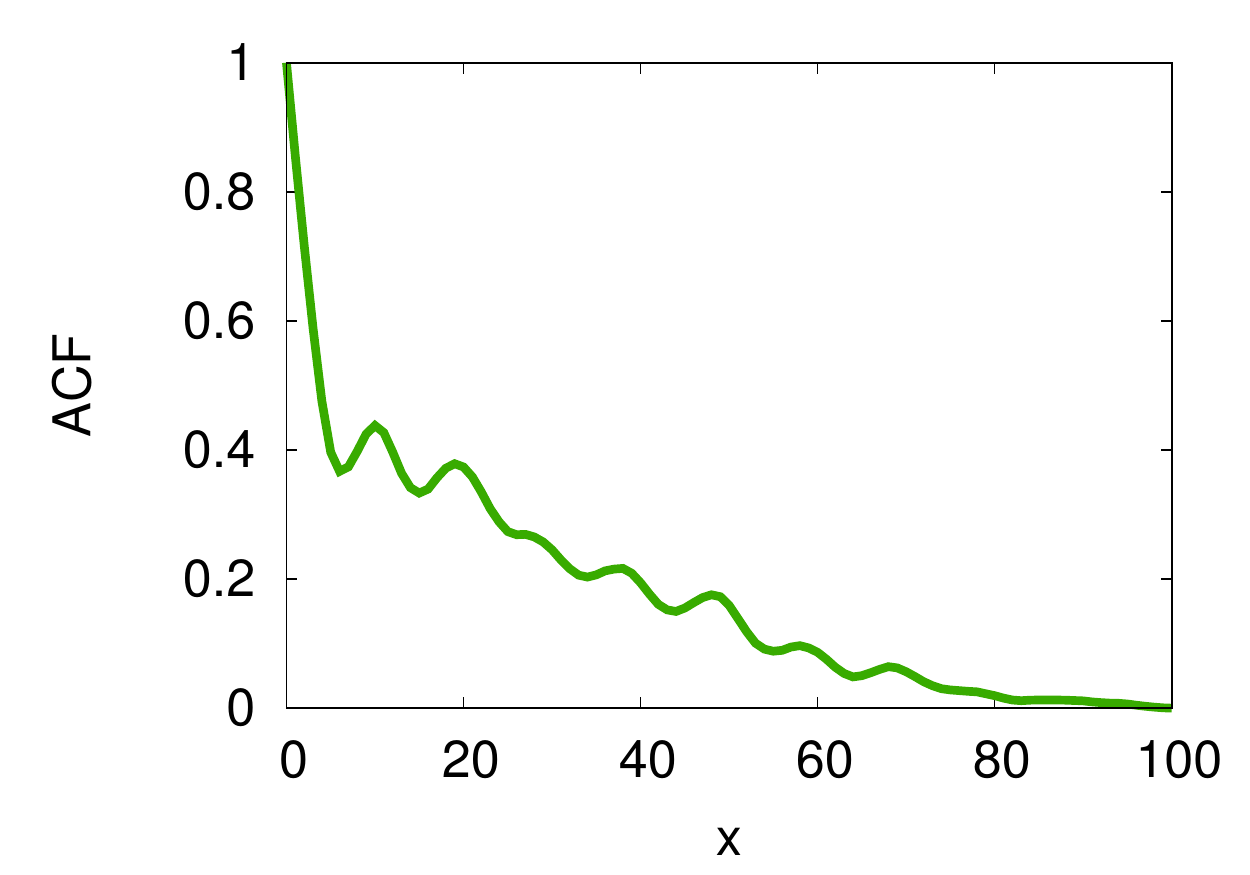} 
\includegraphics[width=0.28\textwidth]{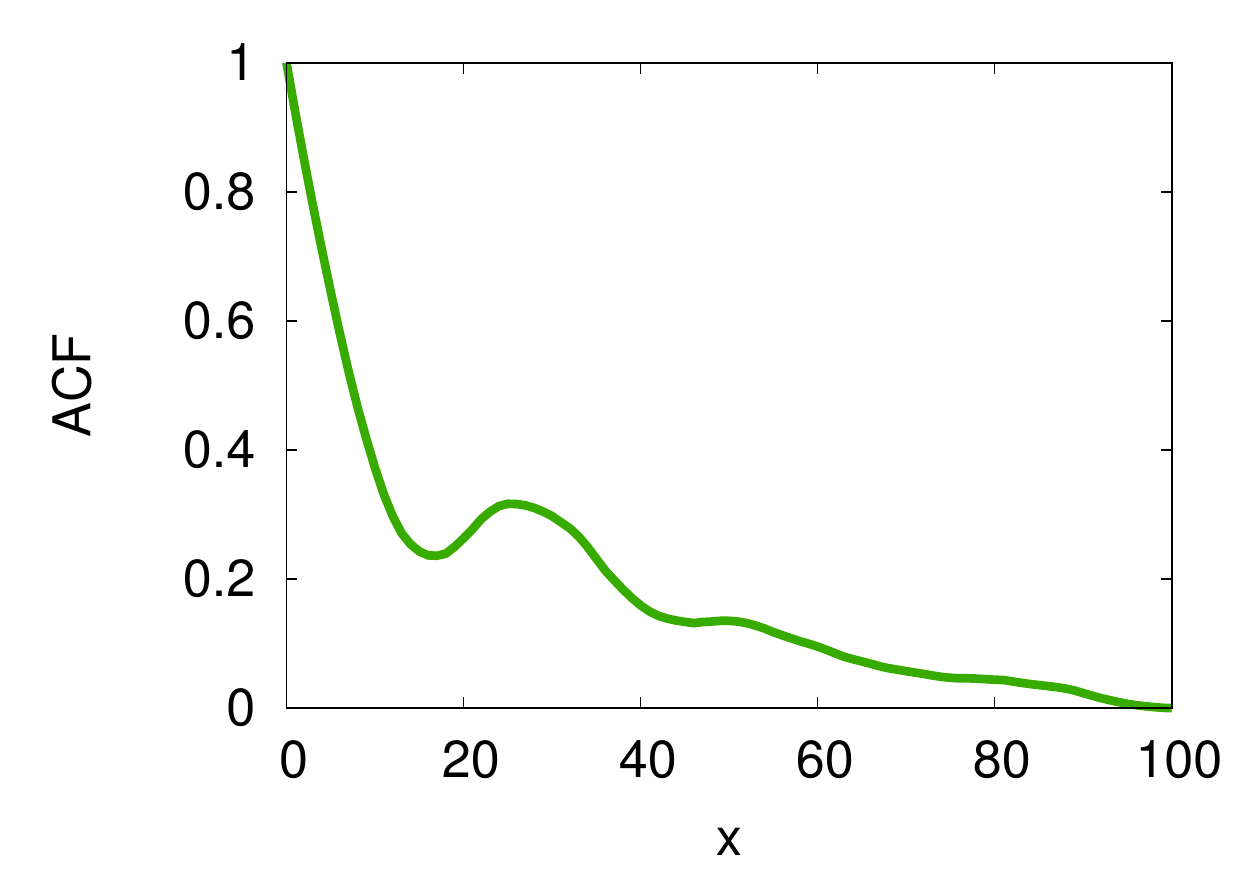} 
\includegraphics[width=0.28\textwidth]{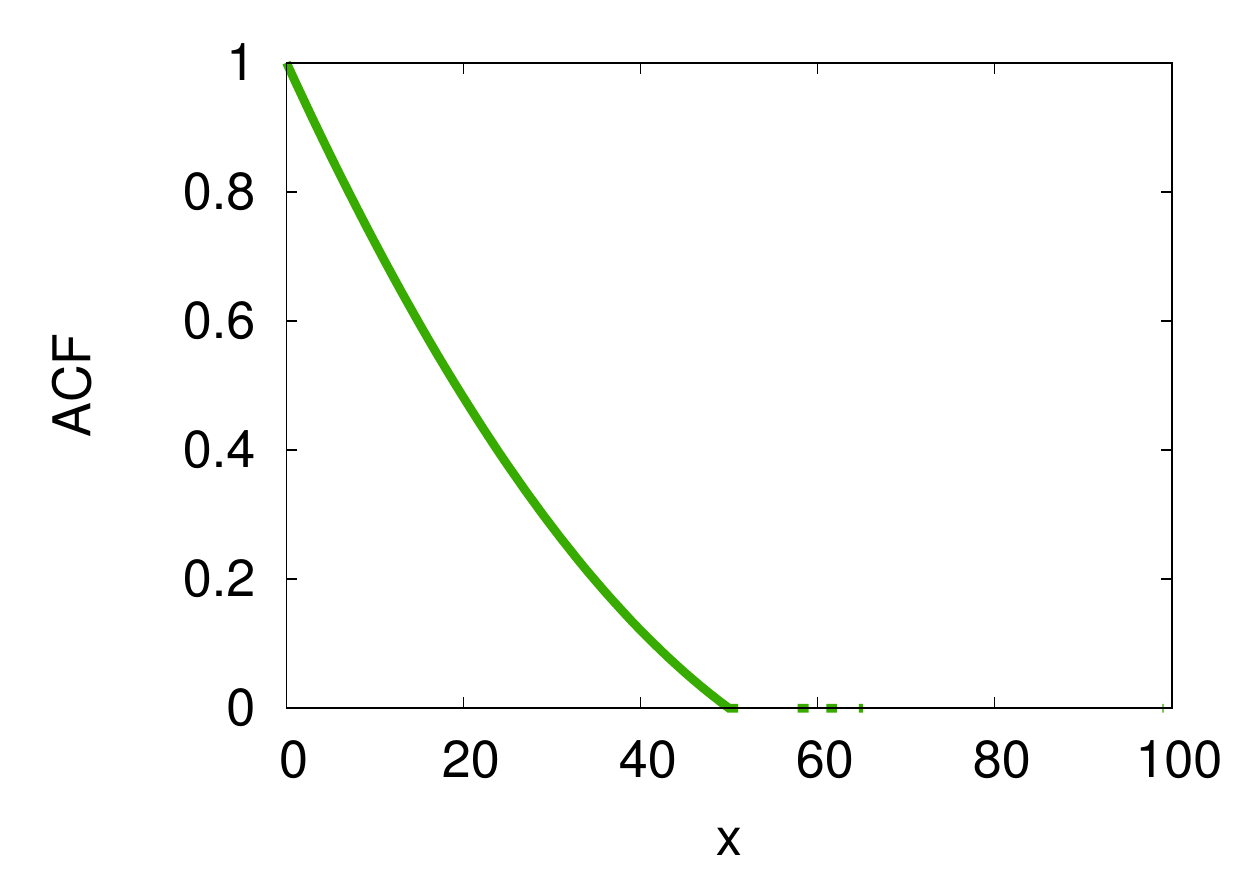} 
\end{center}
\caption{Various types of information extracted from three example morphologies (top). Depicted histograms provide detailed characterization of morphologies with respect to Q2 - what is the distance from any vertex representing photoactive material to the interface? We plot histograms illustrating probability mass function $P(distance=x)$ (second row), and cumulative probability mass function $P(distance<x)$ (third row), and autocorrelation function -- for comparison (bottom).}
\label{fig:results:examples:black:to:green}
\end{figure}

In Figure~\ref{fig:results:examples:black:to:green} we depict three different morphologies with corresponding histograms. These morphologies are selected due to the dramatic variation in feature size and distance to the interface. Note the cumulative distribution functions provide an easy way to estimate the fraction of material less than a specified distance from the interface. 
For comparative purposes, the bottom row of  Figure~\ref{fig:results:examples:black:to:green} plots the autocorrelation function (ACF). The first minimum of the ACF represents the {\it approximate feature size} of the morphology~\cite{MLC09}. The approximate feature size provides a rough estimate of the distance excitons need to diffuse to reach the donor-acceptor interface. However, the ACF cannot provide the \emph{actual fraction} of the domain that can potentially contribute to exciton dissociation (the domain with distance to donor-acceptor interface shorter than exciton diffusion length). 
In contrast, this fraction can be easily read from the cumulative histogram extracted using graph-based analysis (row 3 in Figure~\ref{fig:results:examples:black:to:green}). 
We further note that averaged distance from donor-material to the donor-acceptor interface can also be computed using our graph approach. Thus, our approach provides a more detailed characterization of the morphology than the ACF.

\subsection {Q3: Interfacial area between donor and acceptor regions}
Interfacial area is important in the context of exciton dissociation. Only at the interface can the exciton separate into charges. We include this estimate to demonstrate that different characterizations can be performed easily using the graph approach. 

\begin{figure}
\begin{center}
\begin{scriptsize}
\parbox{0.38\textwidth}{~}
\parbox{0.2\textwidth}{\includegraphics[width=0.15\textwidth]{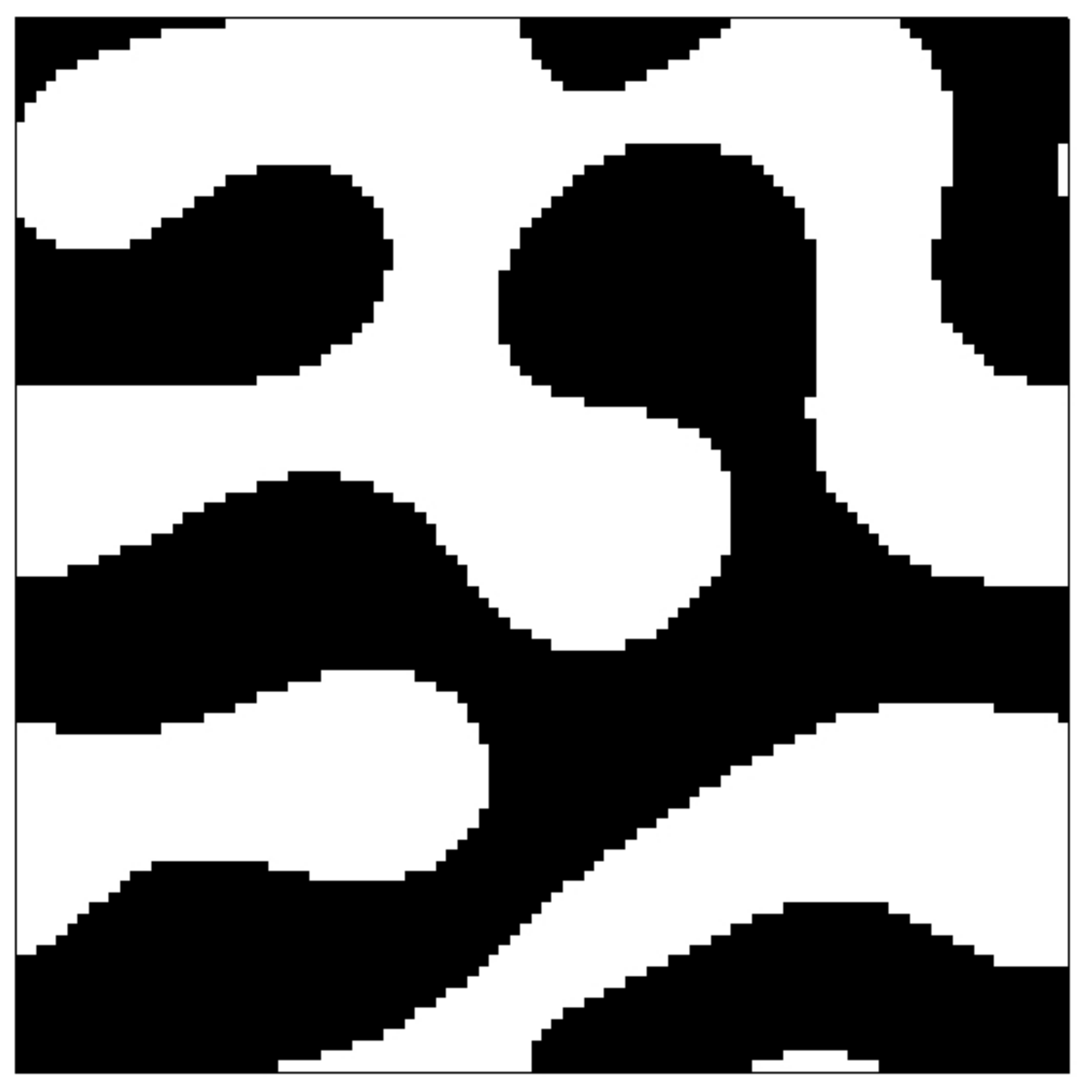} }
\parbox{0.2\textwidth}{\includegraphics[width=0.15\textwidth]{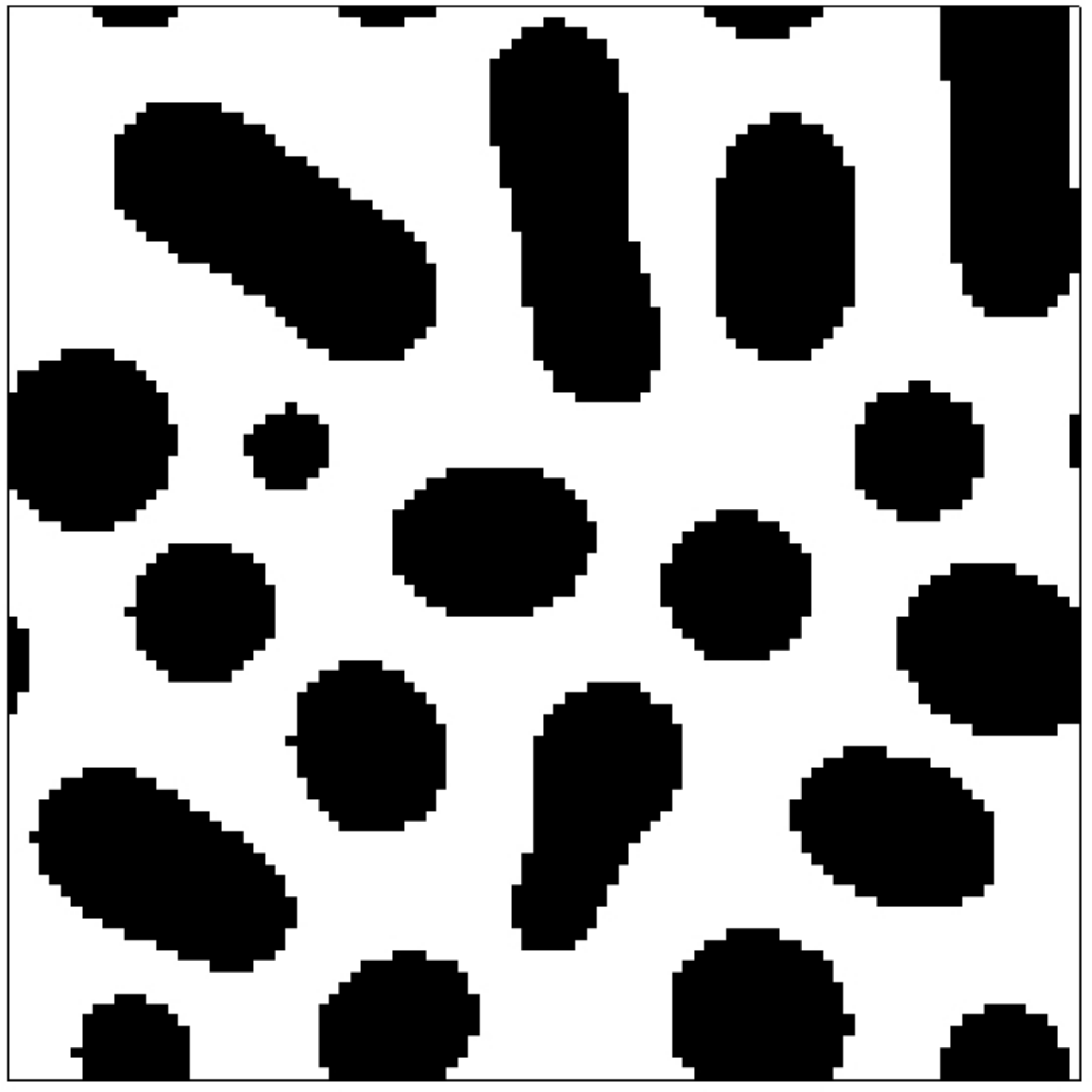} }
\parbox{0.2\textwidth}{\includegraphics[width=0.15\textwidth]{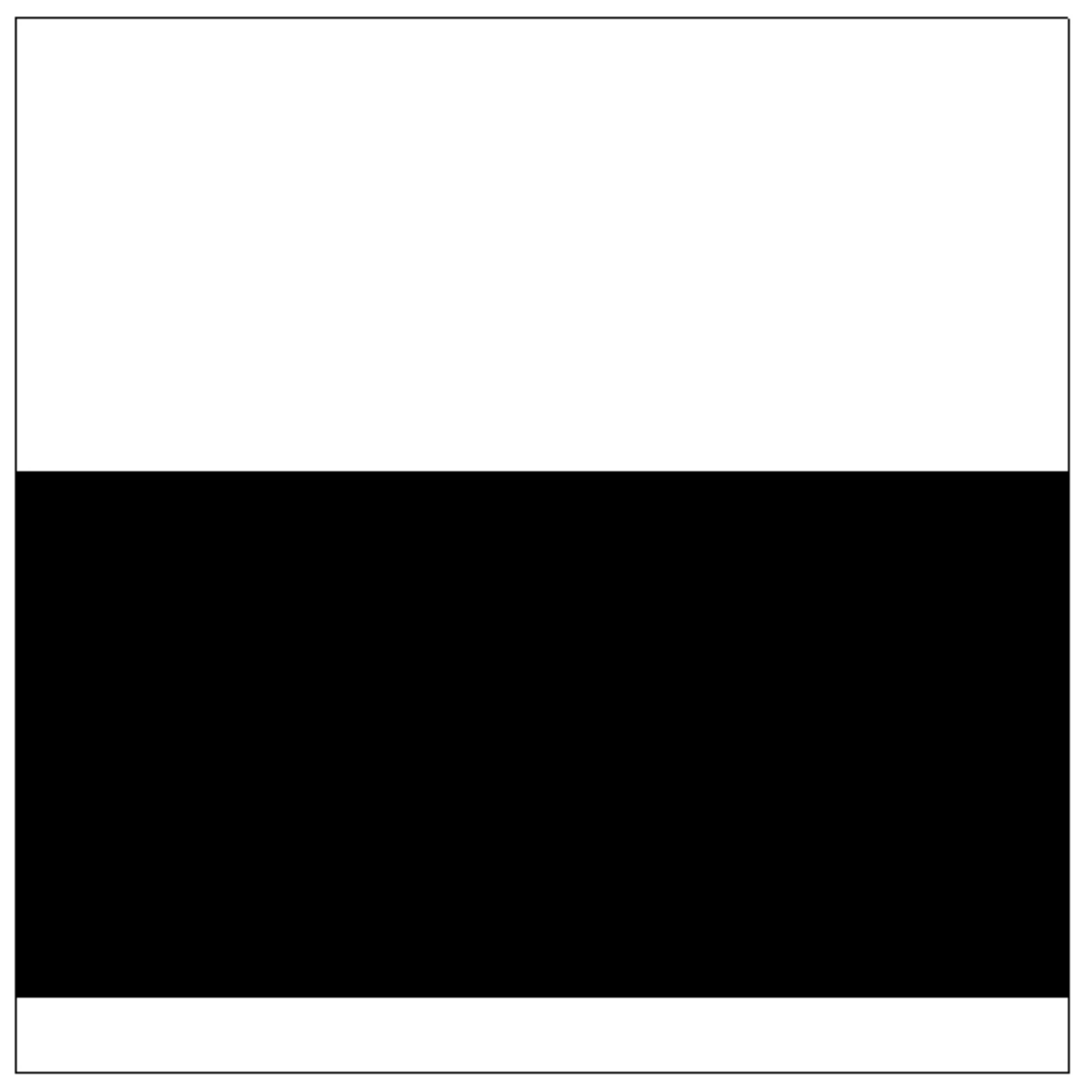} }\\
\parbox{0.38\textwidth}{
Number of black and white components:\\
Number of black components connected to top:\\
Number of white components connected to bottom:\\
Fraction of black vertices connected to top:\\
Fraction of white vertices connected to bottom:\\
Interfacial length (pixels)
}
\parbox{0.05\textwidth}{~}
\parbox{0.1\textwidth}{
6 , 5 \\
3 \\
2\\
12.8\%\\
20.8\%\\
848
}
\parbox{0.1\textwidth}{~}
\parbox{0.1\textwidth}{
24 , 1\\
4\\
1\\
10.3\%\\
100\%\\
1,258
}
\parbox{0.1\textwidth}{~}
\parbox{0.1\textwidth}{
1 , 2\\
0\\
1\\
0\%\\
13.7\%
\\
202
}
\end{scriptsize}
\end{center}
\caption{Three types of morphologies with different coarse grain characterization, domain size is $101\times101$ pixels. }
\label{fig:coarse:info}
\end{figure}

\noindent\framebox[\textwidth]{
	\parbox{0.85\textwidth}{
		\begin{enumerate}
		  \item[Input:] Given labeled, weighted, undirected graph $G=(V,E,W,L)$. 
		  \item Construct set of edges, $I=\{e=(u,v)\in E\;|\; (w(e)=1)\;\land\; ( L(u)=black \land L(v)=white\}$.
		  \item[Output:] $|I|$ -- the cardinality of interface edges. 
		\end{enumerate}
	}
}
The output of the algorithm is the measure of the length (or area) of the interface when all pixels (voxels) are of equal size. We denote the set of interface edges as $I$, and we use this set in subsequent algorithms. Further note an edge $e\in I$ connects black vertex and white vertex with weight $1$.
In Figure~\ref{fig:coarse:info} we present characterization of three morphologies.


\subsection {Q4: Fraction of donor  material connected to anode }

In a BHJ morphology, charges can travel only via one type of material. Holes travel to the anode via electron-donor material, while electrons travel to cathode via electron-acceptor material. A good characterization of charge transport is the {\it fraction of material useful for charge transport}: i.e., the regions which have direct connection to relevant electrode and establish ''pathways'' for charges.\footnote{Islands and domains with no connection to the relevant electrode are not useful for charge transport and can be considered 'dead' regions.}

Determination of the fraction of useful electron-donor material is performed in four steps. We begin by constructing the subgraph induced by black and anode vertices. Next, we compute connected components in the induced subgraph (see Section~\ref{subsec:connComp}). We identify the component $R$ that contains anode in the induced subgraph. The final stage of the procedure is finding the union of the black vertices in component~$R$.

\noindent\framebox[\textwidth]{
	\parbox{0.85\textwidth}{
		\begin{enumerate}
		  \item[Input:] Given labeled, weighted, undirected graph, $G=(V,E,W,L)$. 
		  \item Construct vertex-induced subgraph, $G'=(V',E')$, where $V'=\{v \in V\}$ is a set of all black and anode vertices in $V$, $E'$ is a set of all edges between vertices in $V'$.
		  \item Compute the set of connected components, $C_B$, in $G'$. 
		  \item Identify component $R$ in $C_B$ that contains the anode. 
		  \item Let $P$ be the union of all black vertices in $R$.  
		  \item[Output:] $|P|/|B|$, where $B$ is the set of black vertices.  
		 \end{enumerate}
	}
}
{\bf Remark 2:} To find the fraction of acceptor material connected to the cathode, we design the algorithm by analogy. We proceed by using the same steps as the above algorithm with one difference --  replace the black vertices by white, and the anode by cathode (step 1, step 3, and output). 

\subsection {Q5: Distance from interface to anode via donor domain only}

This question is related to the tortuosity of paths from the interface to the electrodes (see Section~\ref{ch:Mo}). 
This is a more complex computation, but can be easily handled by our framework. 
The graph-based algorithm to compute the quantities is given below:

\noindent\framebox[\textwidth]{
	\parbox{0.85\textwidth}{
		\begin{enumerate}
		  \item[Input:] Given labeled, weighted, undirected graph, $G=(V,E,W,L)$.
		  \item Construct vertex-induced graph, $G'=(V',E')$, where $V'=\{v \in V\}$ is a set of all black and anode vertices in $V$, $E'$ is a set of all edges between vertices in $V'$.
		  \item Let $V_I\subset V'$ be a set of vertices in $V'$ adjacent to interface vertex in $G$.
		  \item Find shortest paths from the anode to all black vertices in $G'$.
		  \item[Output:] Paths from all vertices in $V_I$ to anode and corresponding distances.
		 \end{enumerate}
	}
}
{\bf Remark 3:} To find the distances from the donor-acceptor interface to cathode via acceptor material only, we design the algorithm by analogy. We proceed by using the same steps as the above algorithm with one difference -- replace the black vertices by white, and the anode by cathode (step 1, and 3). 

\begin{figure}
\begin{center}
\includegraphics[width=0.15\textwidth]{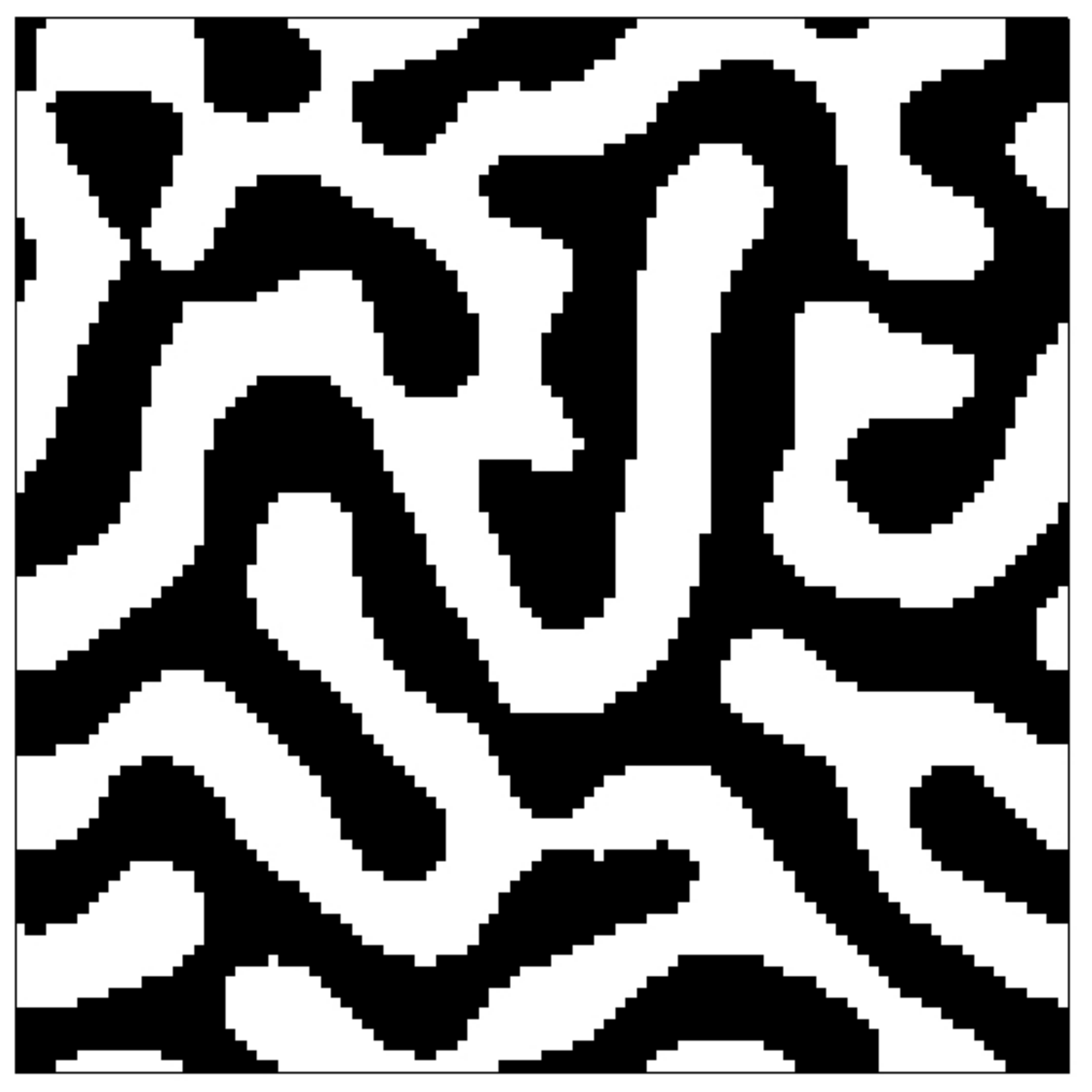} \hspace{0.8cm}
\includegraphics[width=0.15\textwidth]{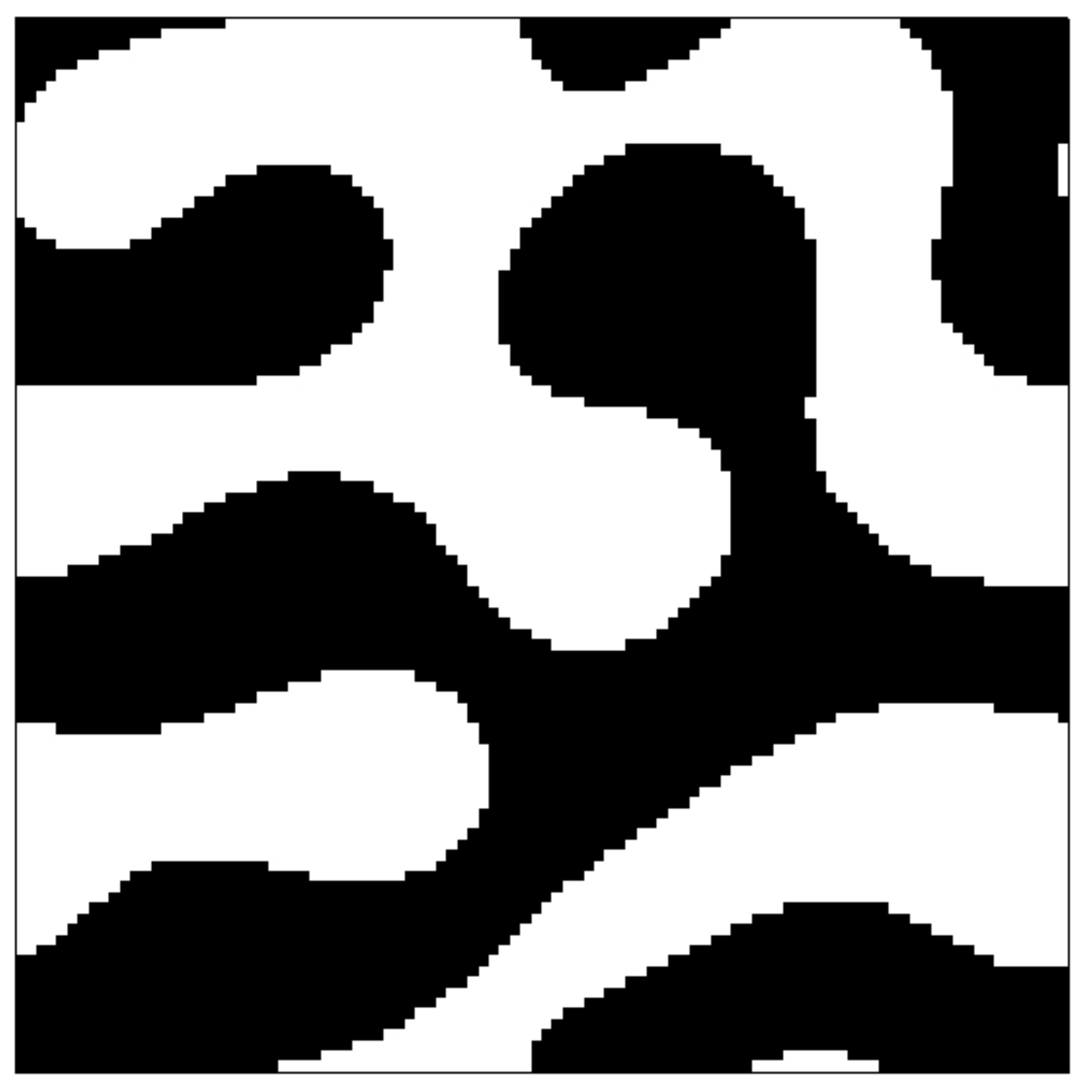} \hspace{0.8cm}
\includegraphics[width=0.15\textwidth]{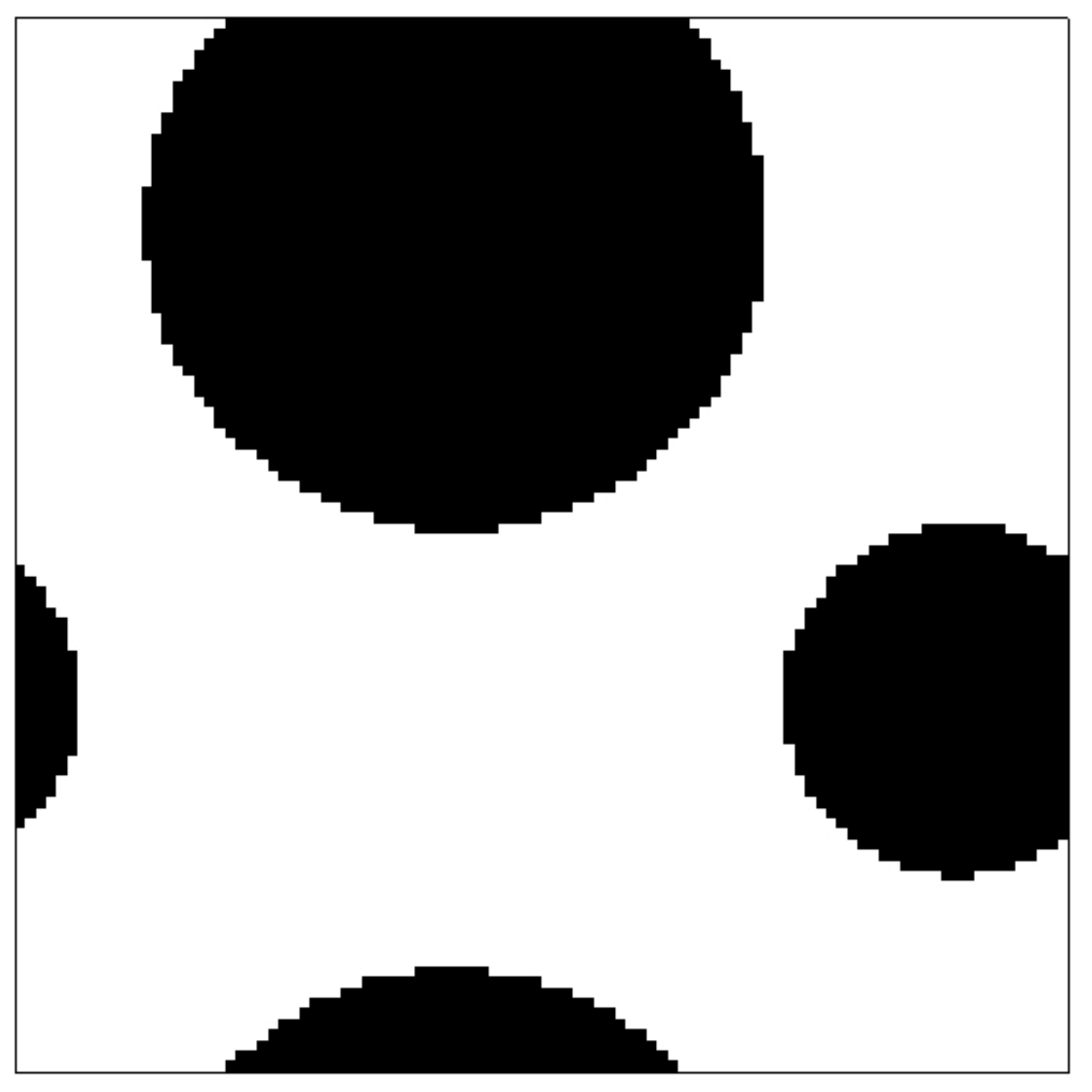}
\\
\includegraphics[width=0.25\textwidth]{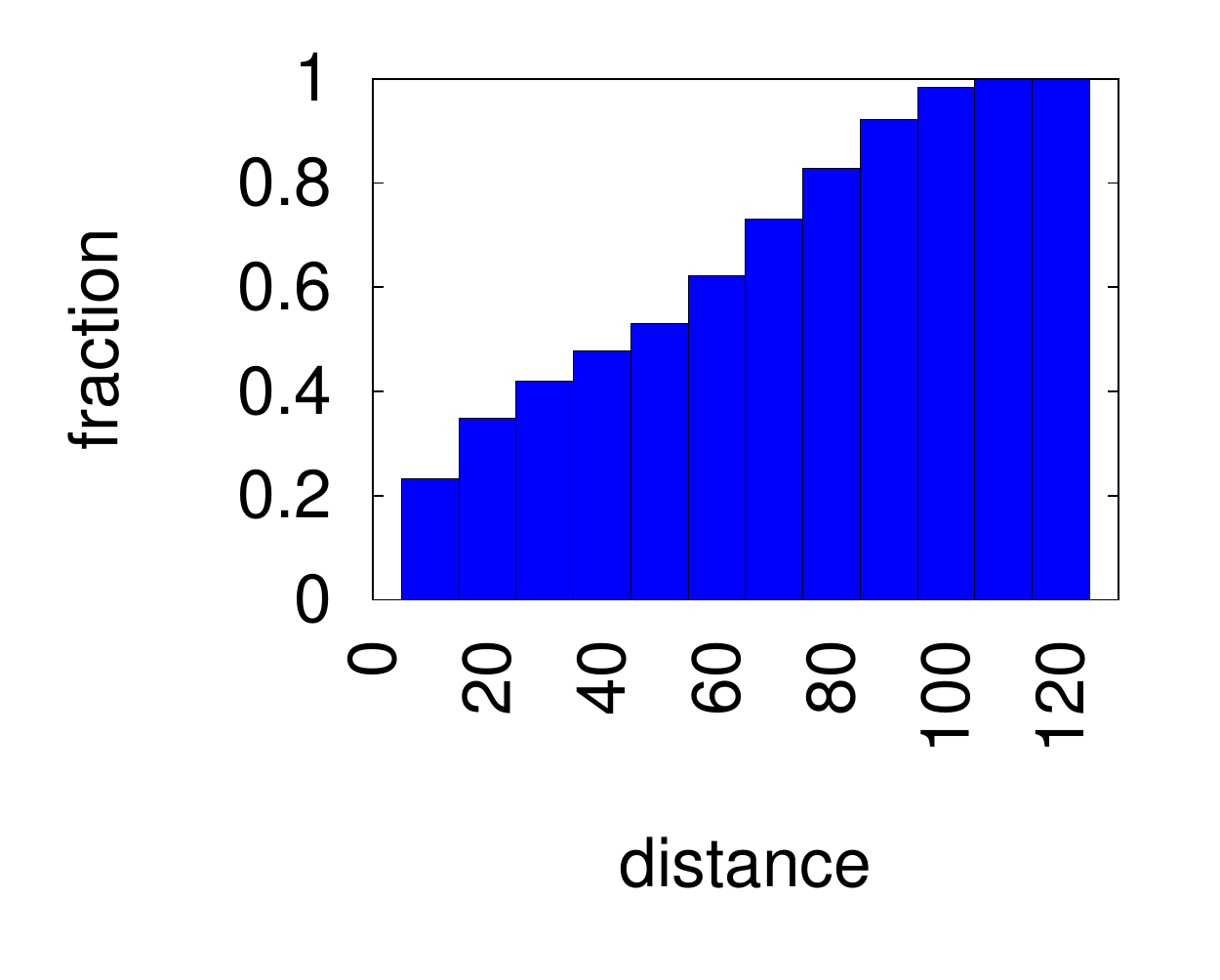} 
\includegraphics[width=0.25\textwidth]{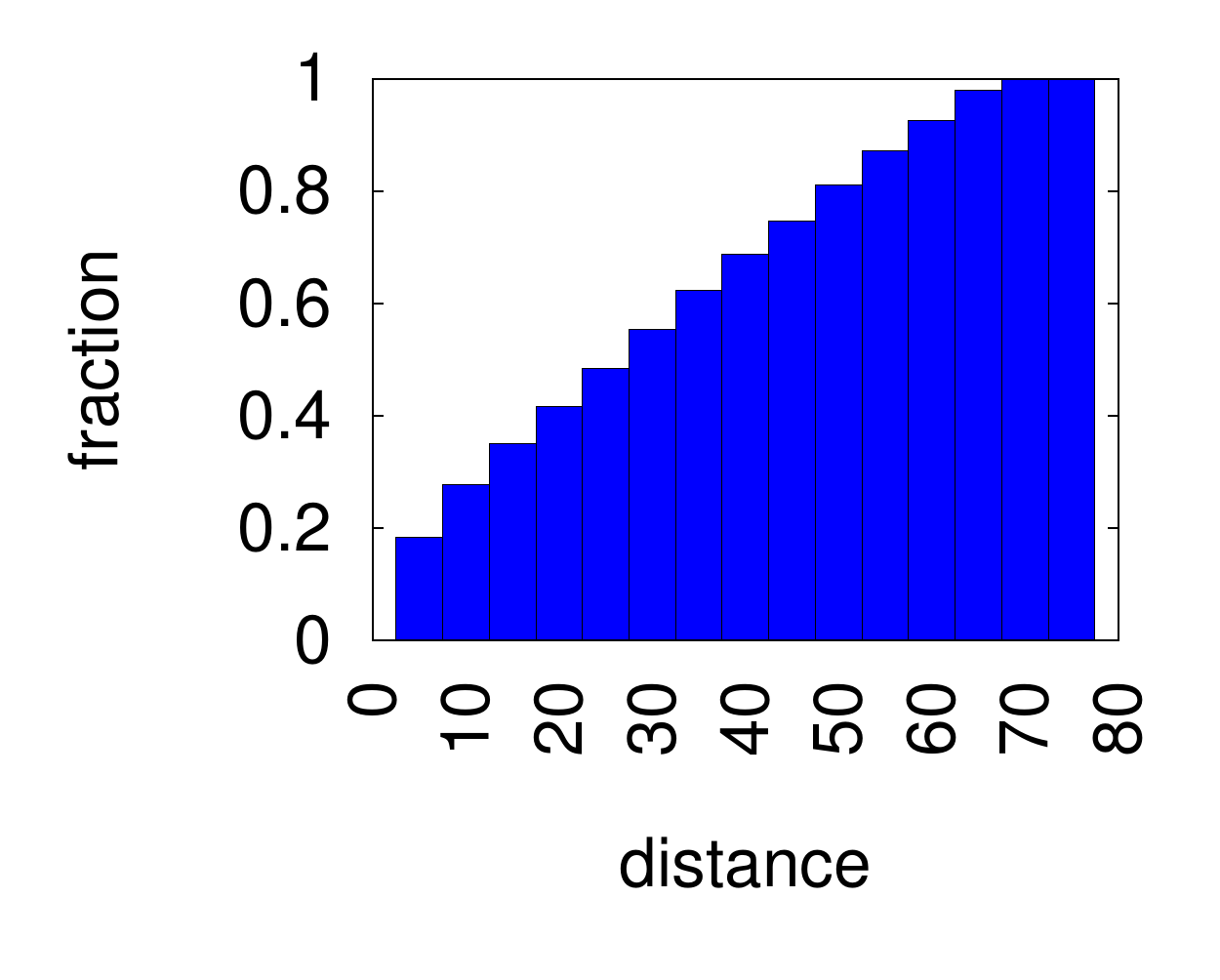} 
\includegraphics[width=0.25\textwidth]{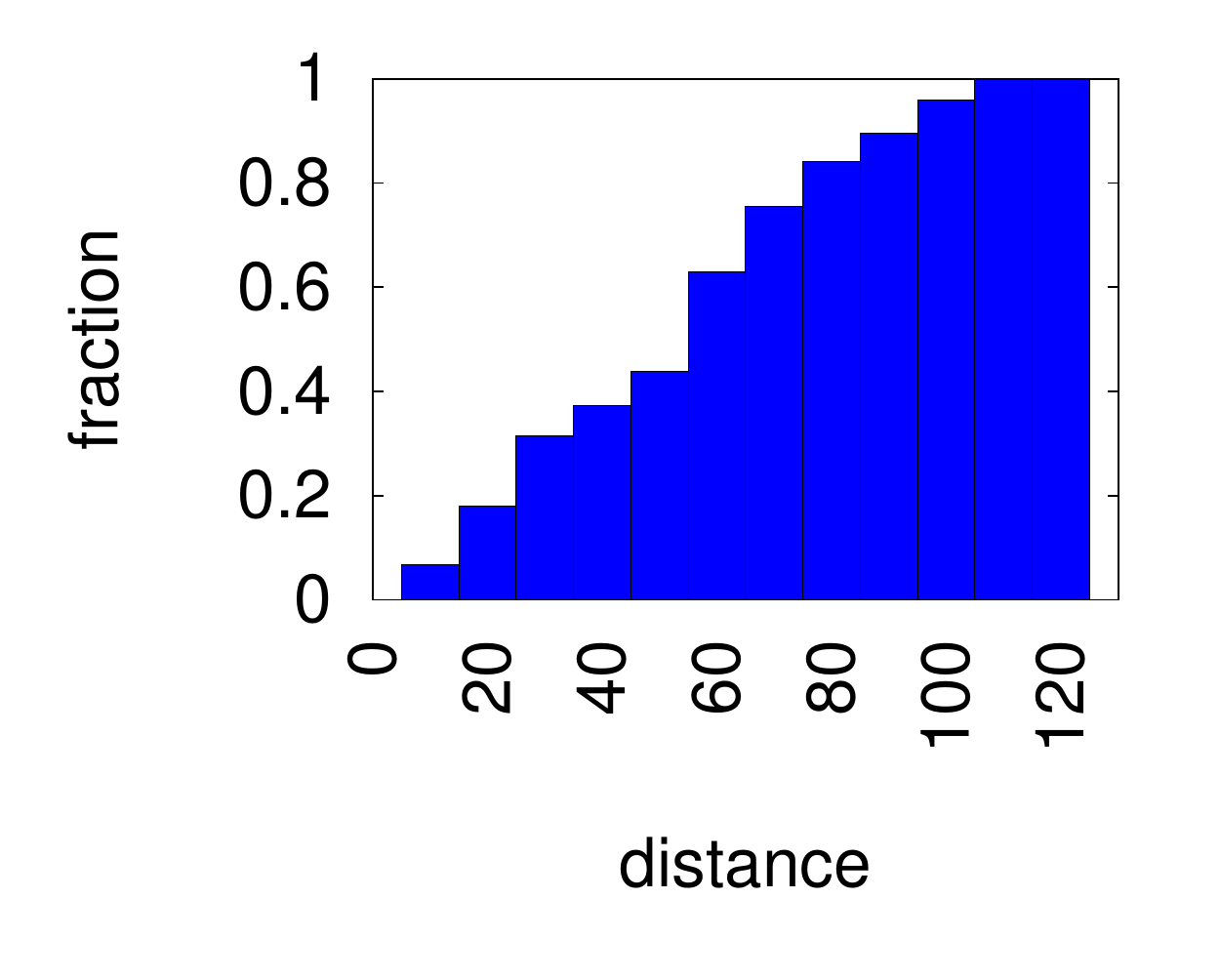} 
\\
\includegraphics[width=0.25\textwidth]{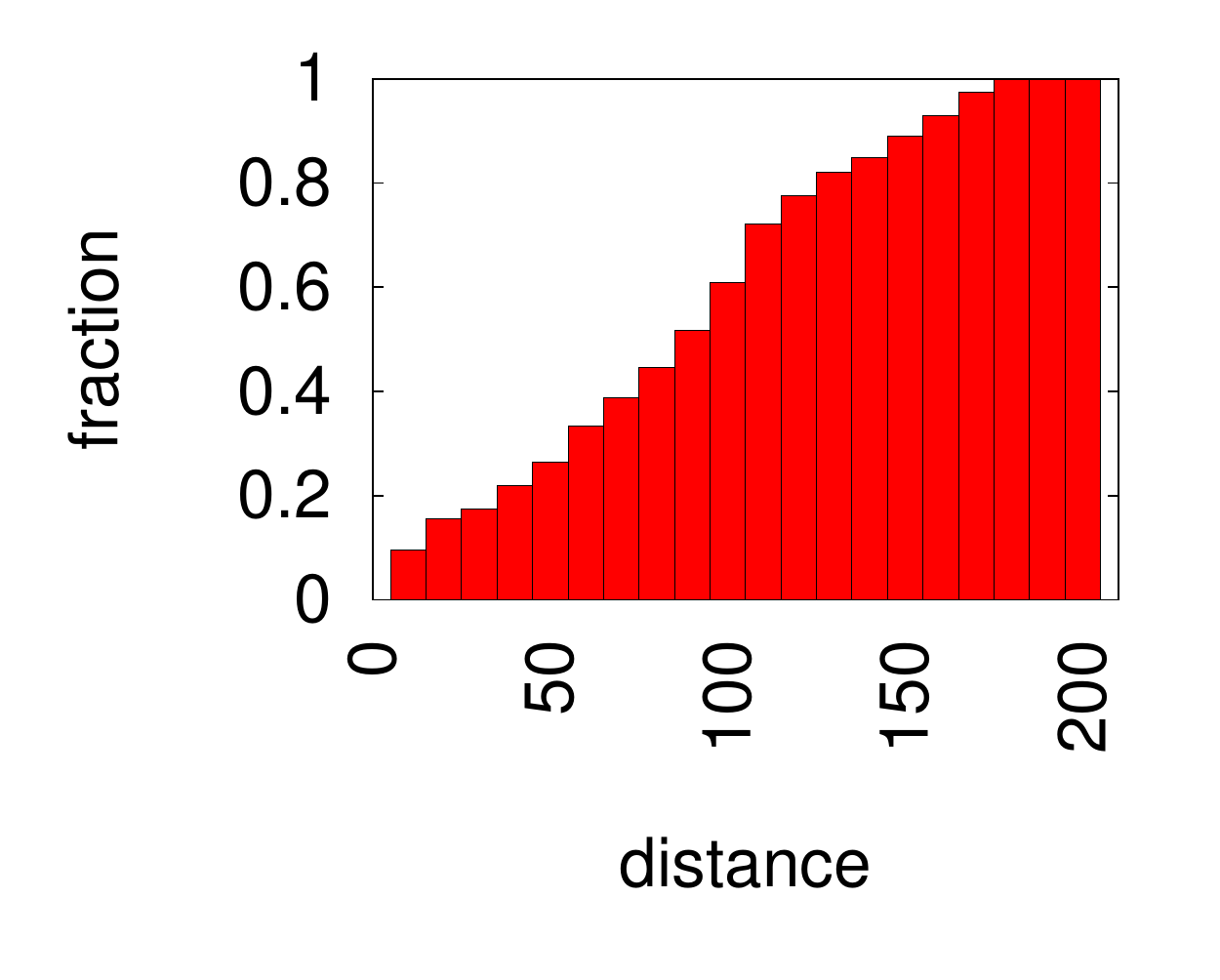} 
\includegraphics[width=0.25\textwidth]{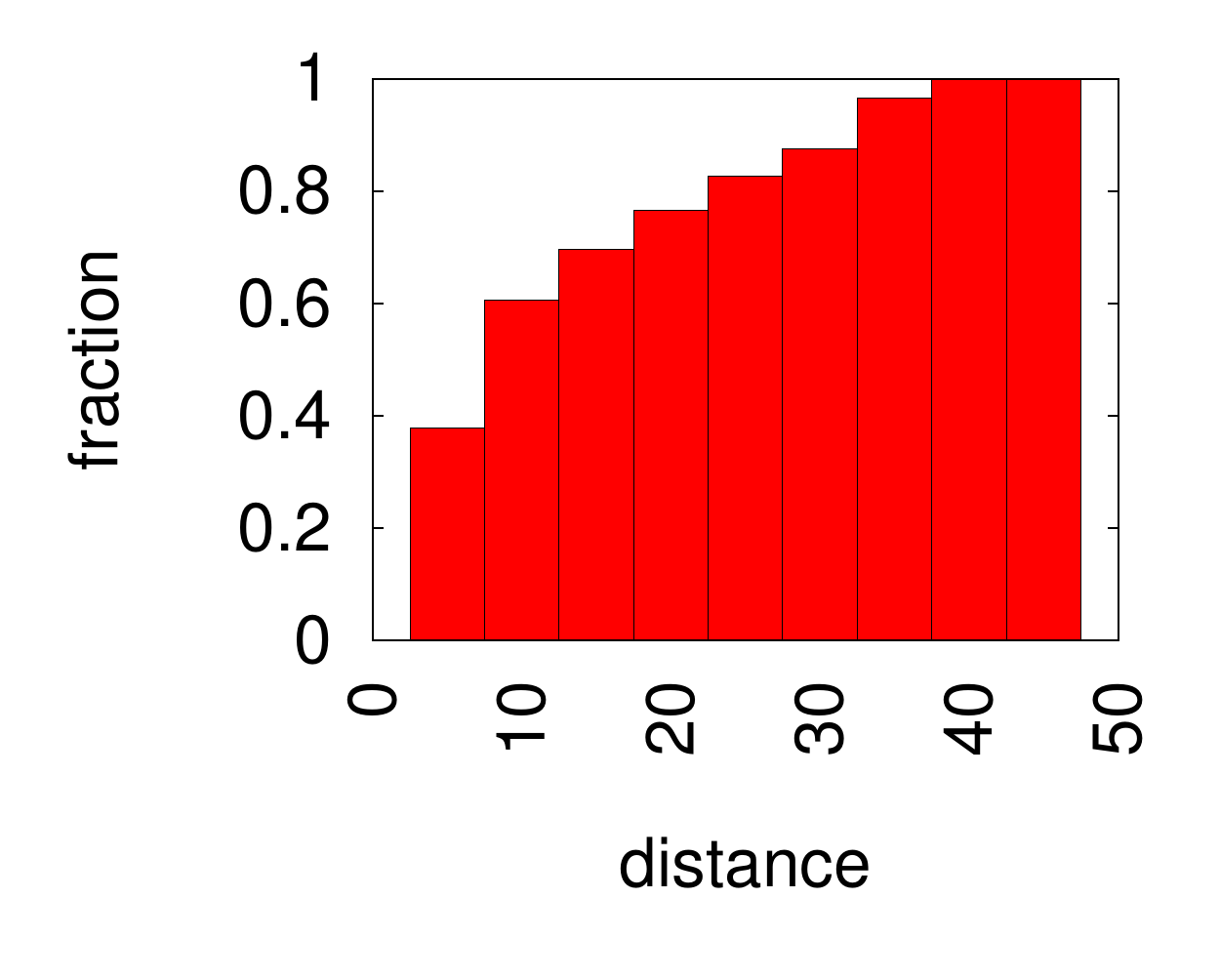} 
\includegraphics[width=0.25\textwidth]{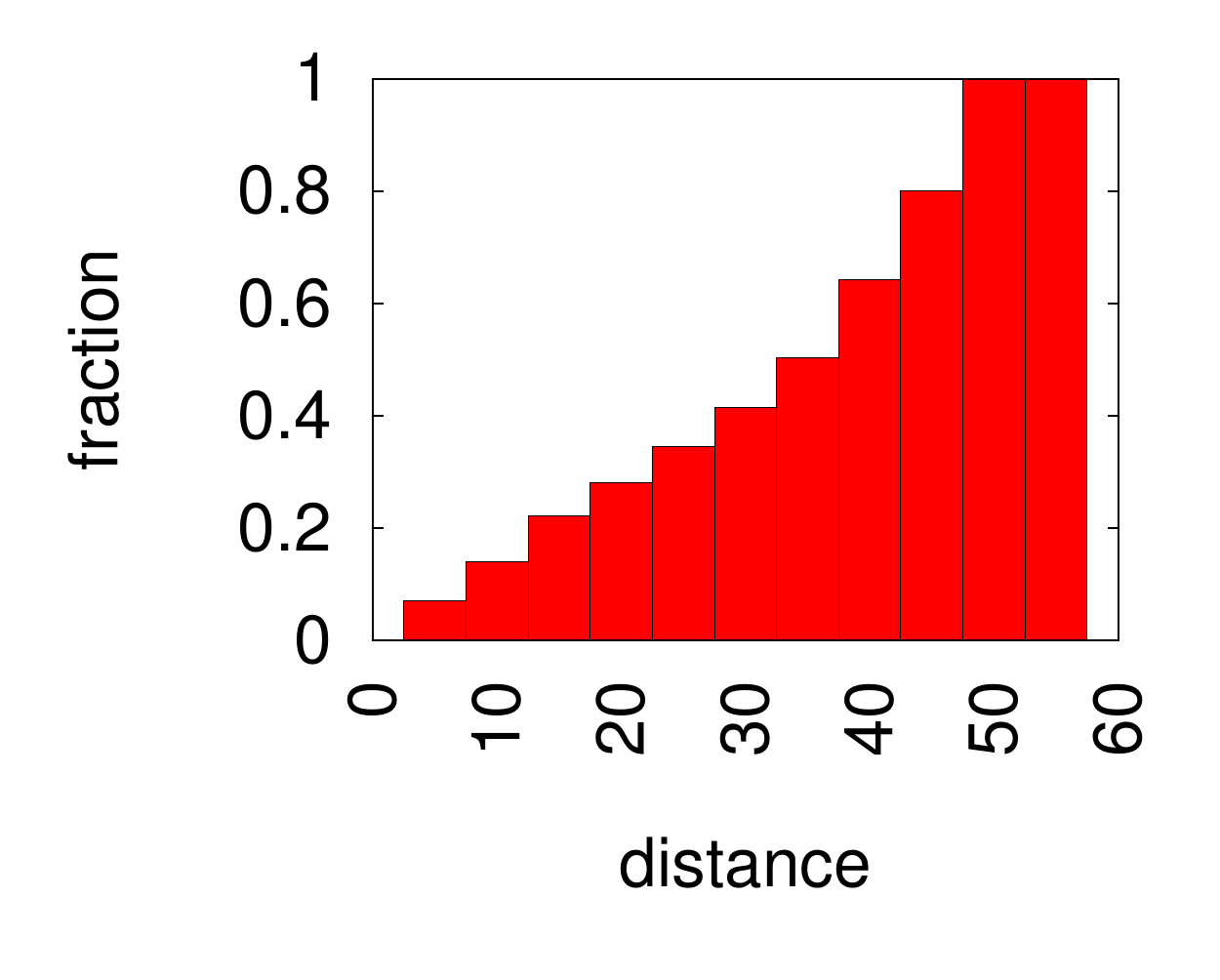} 
\end{center}
\caption{Cumulative histograms of distances from interface to anode via donor (middle row) and from interface to cathode via acceptor (bottom row) extracted for three example morphologies (top row). Histograms provide detailed characterization with respect to Q5.}
\label{fig:results:examples:green:to:red:via:black}
\end{figure}

Figure~\ref{fig:results:examples:green:to:red:via:black} illustrates the above procedure to compute the tortuosity of paths in example morphologies.

\newpage
\subsection {Q6: Interface fraction with complementary paths to both electrodes}

Q4 and Q5 characterize the morphology in terms of useful faction of domain and path lengths for electron and hole to travel. The next level of characterization (as discussed in Section~\ref{ch:Mo}) can be performed by taking into account the fact that charge transport is a complementary process. Specifically, even if one type of charge has perfect pathways to reach the electrode, while its complementary charge does not, the overall transport efficiency deteriorates, due to space-charge effects and recombination. 

Therefore, we search for interface edges in the graph that connect black and white vertices, where both vertices have paths to relevant electrodes (black vertex has path to anode and white vertex has path to cathode) (see Q4). Formally, 
\begin{itemize}
	\item Let $G_B$ be the vertex-induced subgraph with black and anode vertices. 
	\item Let $G_W$ be the vertex-induced subgraph with white and cathode vertices. 
	\item Find a set $I_c\subset I$ of interface edges in the graph, such that the path from the black vertex to anode exists in $G_B$ and the path from white vertex to cathode exists in $G_W$. 
\end{itemize}

This is computed using the following algorithm:

\noindent\framebox[\textwidth]{
	\parbox{0.85\textwidth}{
		\begin{enumerate}
		 \item[Input:] Given labeled, weighted, undirected graph, $G=(V,E,W,L)$. 
		 \item Construct a filtered graph, $G'=(V,E')$, where  E' is a set of edges connecting vertices of the same color.
		 \item Let $I$ be the set of all interface edges (as in Q3).
		 \item Identify connected components, $C$, in $G'$.
		 \item Compute subset $R_r \subset C$ of all components $c\in C$, such that $c$ has black vertices adjacent to anode in G (Q4).
		 \item Compute subset $R_b \subset C$ of all components $c\in C$, such that $c$ has white vertices adjacent to cathode in G (Q4). 
		 \item Identify subset, $I_{rb} \subset I$ where black vertex belongs to  set of vertices in $R_r$ and white vertex belongs to set of vertices in $R_b$.
		 \item[Output:] $|I_{rb}|/|I|$.
		\end{enumerate}
	}	
}
Q6 is formulated in a hierarchical way using Q3 and Q4. This hierarchical construction is one additional advantage of our graph-based approach.

\section{Computing constitutive efficiencies: upper bounds and estimates}
\label{sec:estimates}

The previous section provides a comprehensive and efficient framework to characterize the BHJ morphology through a sequence of formulated questions. We now utilize these characteristics to {\it estimate upper bounds on constitutive efficiencies of a given morphology}. We envision several applications of such estimates, including the search for optimal morphology, and designing fabrication processes which lead to highly efficient devices.  The first application is especially important, since the optimal morphology for OSCs is arguably still unknown. Additionally, designing fabrication processes to obtain highly efficient morphologies is still a complex and challenging task hindered by several factors: (i) lack of tools for systematic search;
(ii) the large phase space for search, and (iii) time consuming simulations 
of device physics to get the final efficiency (e.g., drift-diffusion models~\cite{Koster2005,GommansJanssen2005}, Monte Carlo methods~\cite{MSL10,MarshGreenham2007} are time consuming). 

Our graph-based characterization framework is fast and efficient, and a natural method to search the process space. 
Furthermore, estimating bounds on constitutive efficiencies can be used to filter massive data of morphologies, and identify morphologies with the highest potential for efficient devices. 
Such a rapid prototyping stage for linking structure with properties can significantly speed up the design process for achieving high efficiency morphologies, in contrast to the trial-and-error approach typically followed in morphology optimization experiments.

The reminder of this section details the procedure to compute the upper bounds of each stage of the photovoltaic process, based on the set of questions formulated in Section~\ref{sec:char:BHJ}.

\subsection{Absorption efficiency}

In an OSC, light is typically absorbed only by one type of material - mostly by the electron donating polymer. Thus, the first question from Section~\ref{sec:char:BHJ} (what is the fraction of black vertices?) provides an upper bound on absorption efficiency -- $\eta_{abs}^{upp}$.  This bound, however, is a very conservative bound that does not account for the variation in absorption with depth. 
Intensity of light decays exponentially as light traverses through the active-layer (see Appendix).
Taking this decay into account gives a tighter upper bound on absorption efficiency. The estimated absorption efficiency, $\eta_{abs}^{est}$, is computed by taking into account a weight function assigned to each vertex. The weight function, $a$, assigns value, $a(v)=exp(-\frac{h(v)}{H_d})$, to each vertex of $V$, where $h$ is the physical distance of a vertex, $v$, from the top surface, and $H_d$ is absorption coefficient that is a property of a given material. In this way, instead of counting all vertices (Q1), we count all vertices weighted by their attribute, $a(v)$.  
These are simple estimates which do not take into account the various possible designs to light trapping~\cite{RimPeumans2007}. Nevertheless, they provide important insights about the amount of light that can be absorbed in the device with given morphology. Interference and light trapping effects can be  accounted for by changing the weights of the vertices. 

\subsection{Exciton dissociation efficiency}

We construct two estimates of the exciton dissociation efficiency. The first is based on the maximum possible distance, $L_d$, an exciton can diffuse before it returns to ground state. In practice, $L_d$ is associated with the exciton diffusion length~\cite{Shaw2008}. An upper bound on the exciton diffusion efficiency, $\eta_{diss}^{upp}$, is computed by determining the fraction of the light absorbing domain within a distance, $L_d$, to the donor-acceptor interface. A more aggressive bound, $\eta_{diss}^{est}$, can be computed by picking a smaller  distance, $L_m$, within which the exciton has a {\it high probability of reaching} the interface.\footnote{If $L_m$ is relatively small, i.e., $1\;nm$, then this bound can be associated with the interfacial area.} Both estimates are determined from Q2 defined in Section~\ref{sec:char:BHJ} (what is the fraction of vertices with distance to interface shorter then given distance $d$?). 

\subsection{Charge transport efficiency}

Questions $Q4$ and $Q6$ from the previous section can be used to provide bounds of charge transport. An upper bound for charge transport efficiency can be defined as {\it the fraction of useful regions} for charge transport, $\eta_{out}^{upp}$. Only regions which have a connection to the relevant electrode can contribute to current generation. Thus, computing the fraction of domain satisfying this constraint defines an upper bound, $\eta_{out}^{upp}$. 
This conservative bound can be tightened by considering the complementary nature of the pathways, that is, by computing the fraction of donor-acceptor interface with continuous pathways to both electrodes (Q6), $\eta_{out}^{path}$.
This is because if one type of charge-carriers is not transported to the electrode due to a cul-de-sac, it affects the local electric potential and leads to non-uniform electric field in the device. This, in turn, influences the transport of the other type of charge carriers and increases charge recombination.

\section{Technical Details}
The complete framework is implemented in C++ using the boost library~\cite{Boost}. This library utilizes highly optimized implementations of the two basic algorithms used: shortest path and connected components. The computational complexity depends on the type of data, e.g., number of vertices and edges. The run time depends also on the sparsity of the graph. Dijkstra's algorithm has complexity $\mathcal{O}(n\log(n))$, where $n$ is number of vertices. The complexity of the connected components algorithm is linear, $\mathcal{O}(n)$. Run time for data sets of size $1001\times101$ is, on average, 20 s on a QuadCore 3GHz computer with 8GB of RAM. Scaling analysis of the framework together with implementation details is the subject of a subsequent paper~\cite{WTCG11c}.

\section{Application to analysis of BHJ morphology}

Morphology descriptors formulated in the above sections can be evaluated for any two-phase morphology. Input morphology can either be from experimental observation~\cite{AHM09,BSWL09,MLC09} or numerical simulations~\cite{TsigeGrest2005,PuemansForrest2003,WG10a}. We demonstrate how the constructed suite of morphology descriptors can be employed to characterize experimentally acquired morphology. Subsequently, we exploit the framework to gain insight into the thermal annealing process in OSCs.

\subsection{Characterization of BHJ}
A TEM cross-section image of BHJ taken from Moon et al.~\cite{MLC09} is characterized. Each pixel of the image is represented as a vertex in a graph. The analyzed image consists of $1157\times106$ pixels. 
We showcase the exhaustive characterization of BHJ with respect to all the subprocesses of the photovoltaic effect by answering the six questions formulated in Section~\ref{sec:char:BHJ}. 

We begin by quantifying morphology with respect to light absorption. 
The light-absorbing material occupies 58.3\% area in the shown morphology (Q1).
Thus, this quantity defines the upper bound on absorption efficiency. Additionally, by considering light intensity decrease due to absorption depth, the absorption efficiency can be estimated to be 36\% (see Section~\ref{sec:estimates}).

Next, we proceed to exciton dissociation. Question Q2 allows us to determine the fraction of donor material within a distance, $d$, to the interface. Choosing the distance, $d$, to be the exciton diffusion length ($L_d=10\;nm$) gives us an upper bound on the exciton dissociation efficiency. This reveals that 99\% of the photoactive material is within distance $d<10\;nm$. Furthermore, 21\% of the photoactive material is located within $1\;nm$ to the interface. Consequently, we foresee that almost all excitons are able to reach the donor-acceptor interface, and 21\% of these have a very high chance to reach interface. Both estimates provide a good prognosis for a high efficiency device. 

Finally, we characterize charge transport using a hierarchy of three descriptors: 
\begin{itemize}
	\item[(i)] by determining fraction of useful domains (Q4), 
	\item[(ii)] by determining the fraction of interface with complementary pathways to both electrodes (Q6) and 
	\item[(iii)] by estimating how well-balanced the structure is with respect to charge transport (Q5). 
\end{itemize}
Q4 provides an upper bound on charge transport by determining the fraction of useful domains to be $\eta_{out}^{upp}=79\%$ of overall morphology. That is, $21\%$ of the active layer cannot contribute to charge transport, because there is no direct connection between these regions and relevant electrodes. Separately analyzing donor and acceptor regions, we learn that 97\% of the donor is connected to the anode but only 54\% of the acceptor is connected to the cathode (Q4a, Q4b). In this way, we identify the main bottleneck to charge transport in this morphology. A similar estimate is extracted from Q6: only 47\% of the interface has complementary paths to the respective electrodes. 

Finally, we can estimate the average distances the hole and the electron need to travel to reach the cathode and the anode, respectively. For this morphology, the hole needs to travel an average distance of $67\;nm$ to reach anode. In comparison, the electron needs to travel an average distance of $53\;nm$ to reach the cathode. The ratio between the average distances tells us how well balanced is the structure.\footnote{The structure balance depends also on the charge mobilities, which, together with distances, affect the time charges required to reach the electrode.}

\begin{figure}
\parbox{0.05\textwidth}{~}
\parbox{0.94\textwidth}{\includegraphics[width=0.89\textwidth]{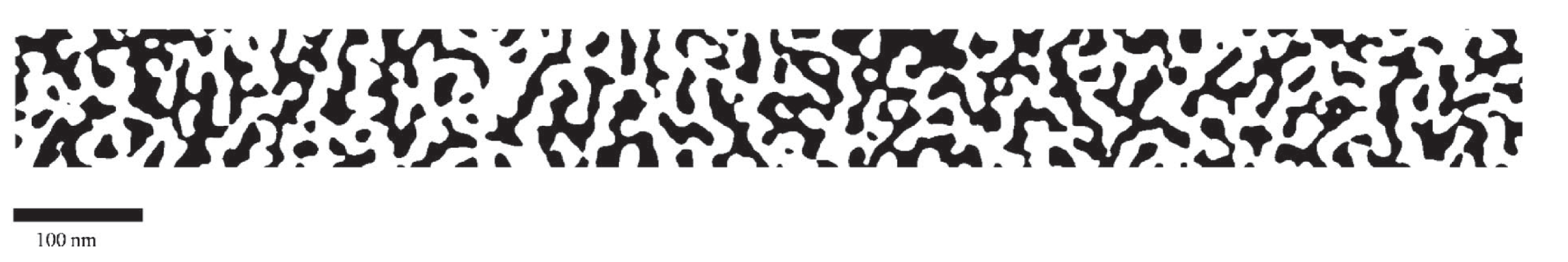}}\\
\parbox{0.05\textwidth}{Q1}
\parbox{0.94\textwidth}{Fraction of light absorbing material: 58.3\%}\\
\parbox{0.05\textwidth}{Q2}
\parbox{0.94\textwidth}{Fraction of photoactive material whose distance to the interface is within exciton diffusion length ($d<10\;nm$): 99.4\%

}\\
\parbox{0.05\textwidth}{Q3}
\parbox{0.94\textwidth}{Interfacial area between donor and acceptor regions? $15,044nm$}\\
\parbox{0.05\textwidth}{Q4a}
\parbox{0.94\textwidth}{Fraction of donor domains connected to the anode:  97\%}\\
\parbox{0.05\textwidth}{Q4b}
\parbox{0.94\textwidth}{Fraction of acceptor domains connected to the cathode: 54\% }\\
\parbox{0.05\textwidth}{Q5a}
\parbox{0.94\textwidth}{Averaged distance from interface to the anode via donor domain:  $67.82\;nm$}\\
\parbox{0.05\textwidth}{Q5b}
\parbox{0.94\textwidth}{Averaged distance from interface to the cathode via acceptor domain: $53.61\;nm$}\\
\parbox{0.05\textwidth}{Q6}
\parbox{0.94\textwidth}{The interface fraction which has complementary paths to both electrodes: 47\%}
\caption{Example two-phase morphology filtered from TEM image of BHJ~\cite{MLC09} with answers to six important questions.}
\label{fig:results:MLC09:120}
\end{figure}

\newpage
\subsection{Analysis of the fabrication process}
We showcase how the estimated efficiency and bounds may be used to identify the optimal morphology. Specifically, we are interested in a rapid investigation of multiple morphologies and selecting the best morphology and its associated fabrication conditions.

\begin{figure}
\begin{center}
\parbox{0.49\textwidth}{
\begin{center}
$1:1$\\
\includegraphics[width=0.45\textwidth]{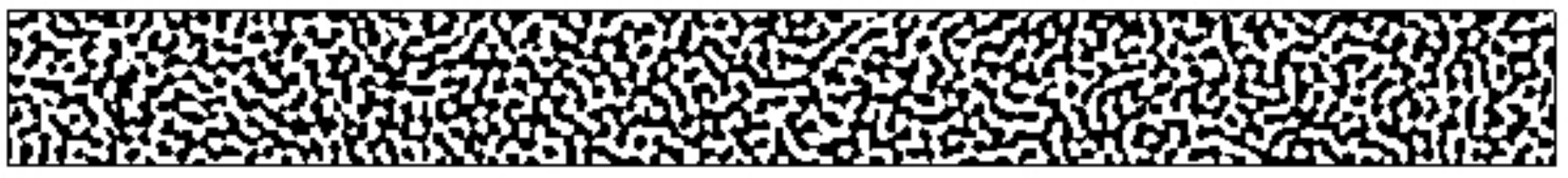}\\
\includegraphics[width=0.45\textwidth]{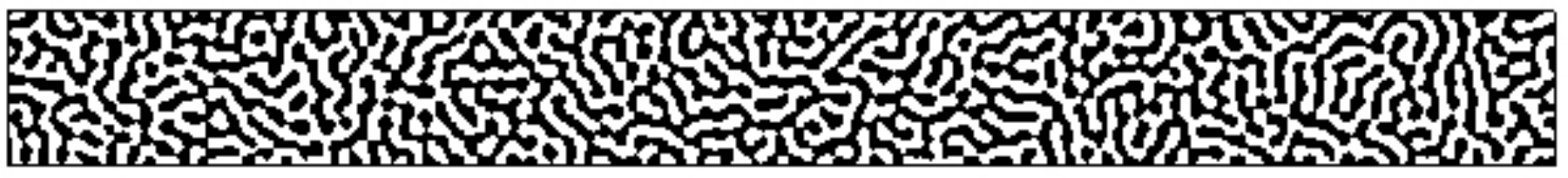}\\
\includegraphics[width=0.45\textwidth]{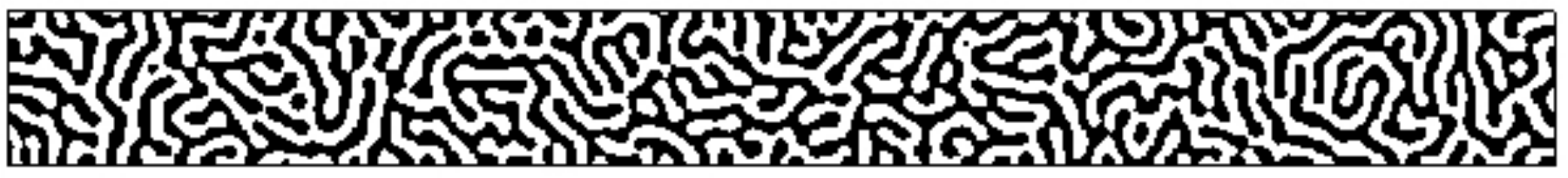}\\
\includegraphics[width=0.45\textwidth]{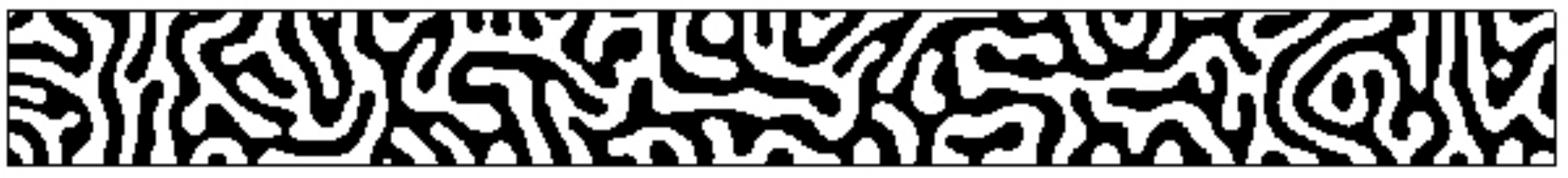}\\
\includegraphics[width=0.45\textwidth]{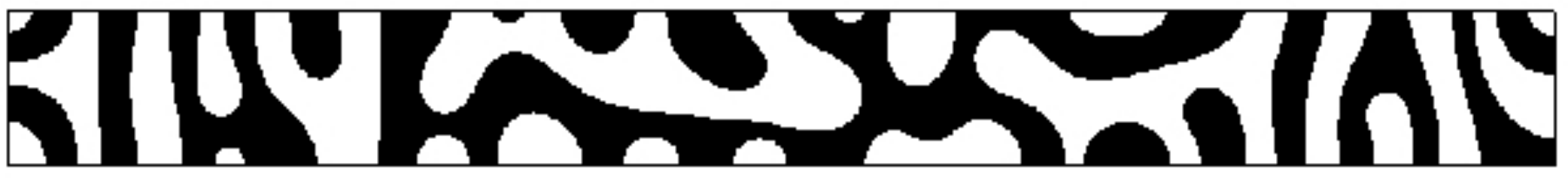}\\
\includegraphics[width=0.45\textwidth]{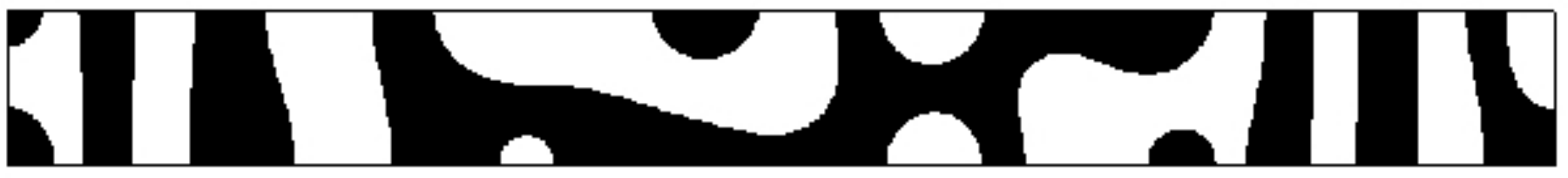}
\end{center}
}
\parbox{0.49\textwidth}{
\begin{center}
$1:0.82$\\
\includegraphics[width=0.45\textwidth]{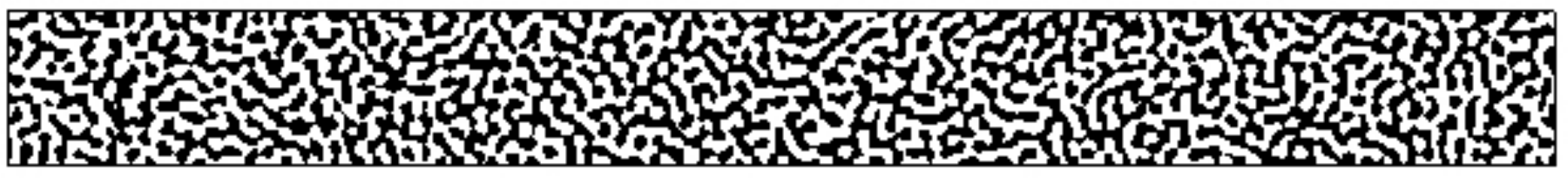}\\
\includegraphics[width=0.45\textwidth]{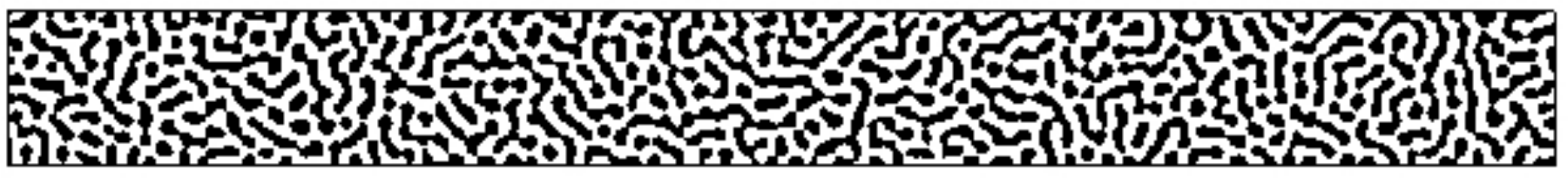}\\
\includegraphics[width=0.45\textwidth]{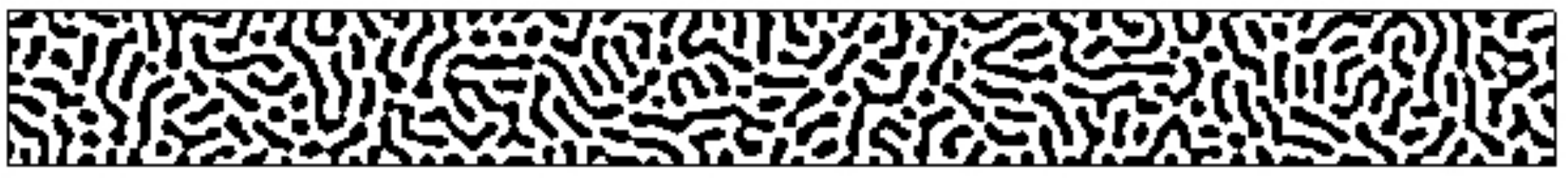}\\
\includegraphics[width=0.45\textwidth]{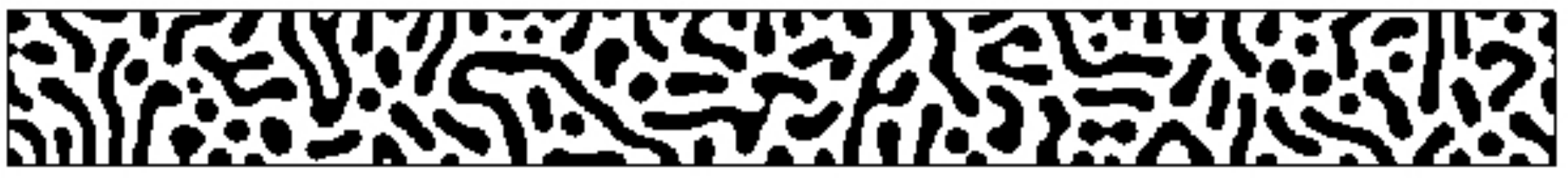}\\
\includegraphics[width=0.45\textwidth]{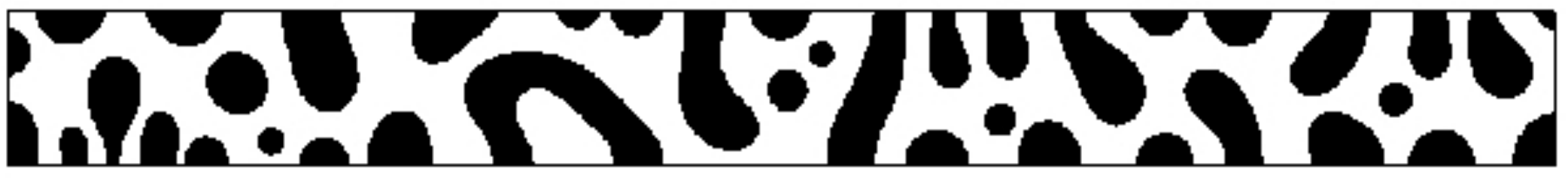}\\
\includegraphics[width=0.45\textwidth]{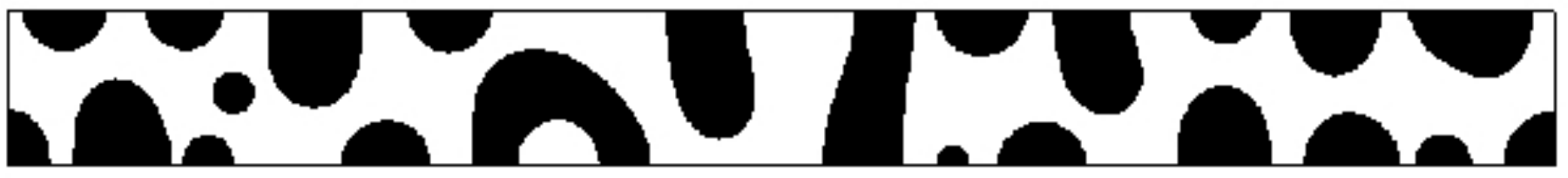}
\end{center}
}
\end{center}
\caption{Representative two-phase morphology evolution obtained via numerical simulation of thermal annealing for two blend ratios (1:1 and 1:0.82) from early stages (first row) untill the final time (consecutive rows). 
}
\label{fig:results:CHeq}
\end{figure}

We start with two sets of morphologies fabricated using blend ratios of 1:1 and 1:0.82 (acceptor:donor), respectively (Figure~\ref{fig:results:CHeq}). Each set of morphologies depicts time evolution during the process of thermal annealing of this model system. Thermal annealing is one of the important post-fabrication processes utilized in OSCs to improve power conversion efficiencies\cite{CHGY09,MouleMeerholz2008,MYG05}.  

We choose these two blend ratios because they represent two types of morphologies --- percolated morphology and morphology with islands. Intuitively, the former is more suitable for OSC than the latter, for all times during annealing. 
This is because the latter structure has more islands not connected to relevant electrodes and cannot provide useful pathways for charges. 
Each set consists of nine hundred morphologies, each represents morphology of the size 100 nm$\times$1000 nm. Figure~\ref{fig:results:CHeq} depicts six representative morphologies for each type. These morphologies are obtained by solving the Cahn-Hilliard equation (for more details refer to~\cite{WG11a}). 
We utilize a graph-based framework to quantify the differences between two morphologies, thus, validating this intuitive hypothesis.

\begin{figure}
\parbox{0.5\textwidth}{(a)}
\parbox{0.5\textwidth}{(b)}\\
\includegraphics[width=0.49\textwidth]{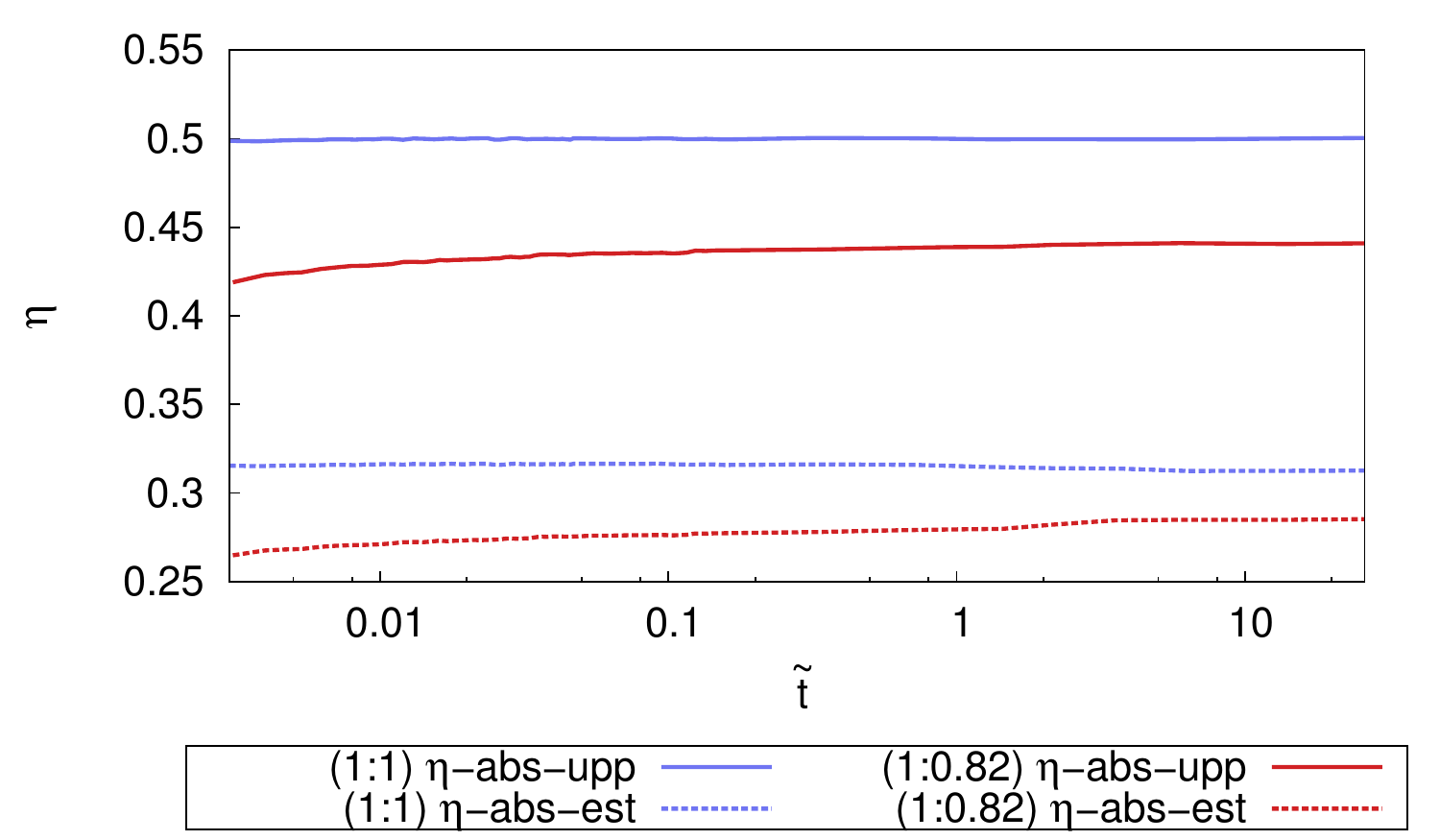}
\includegraphics[width=0.49\textwidth]{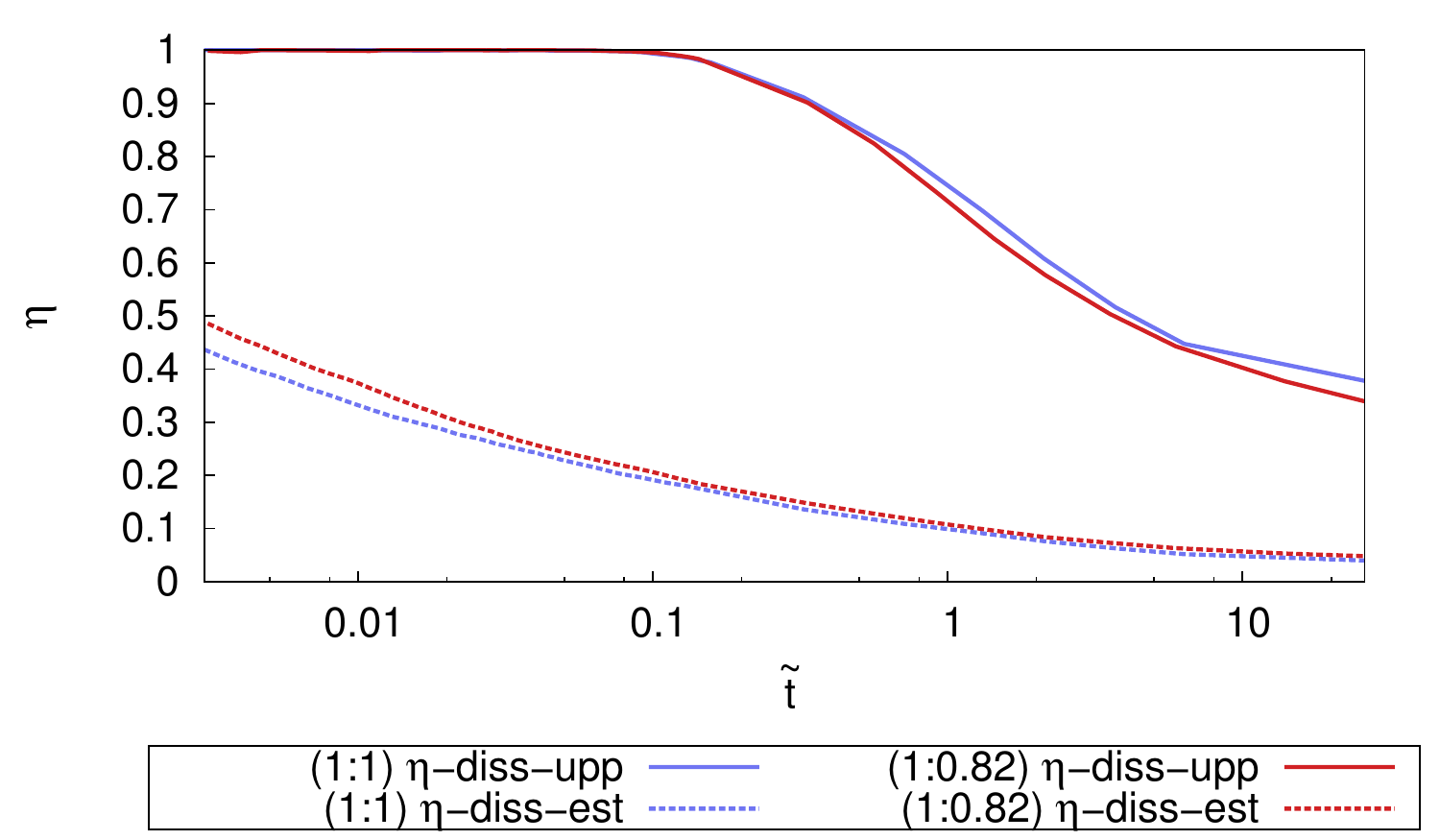}\\
\parbox{0.5\textwidth}{(c)}
\parbox{0.5\textwidth}{(d)}\\
\includegraphics[width=0.49\textwidth]{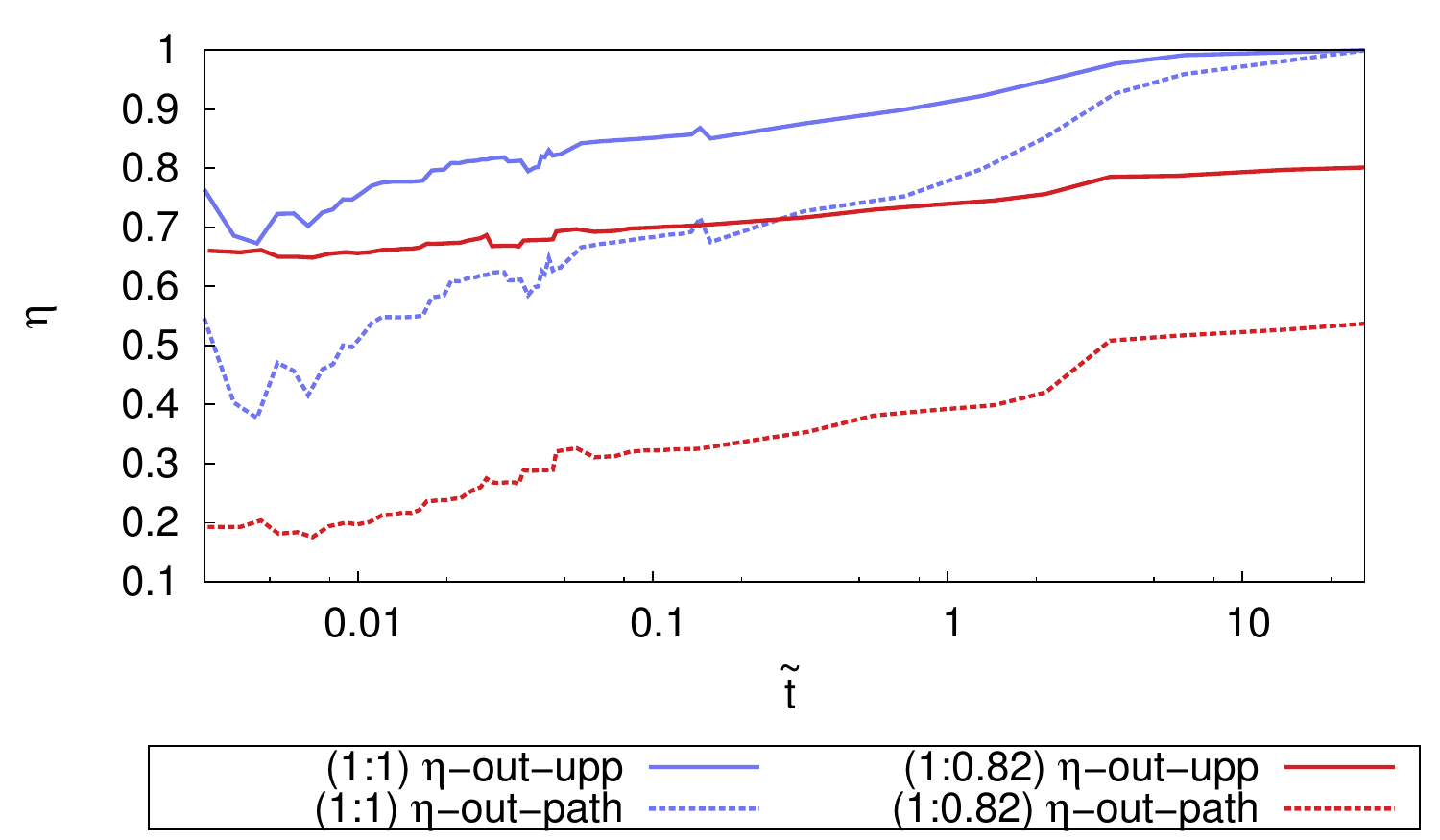}
\includegraphics[width=0.49\textwidth]{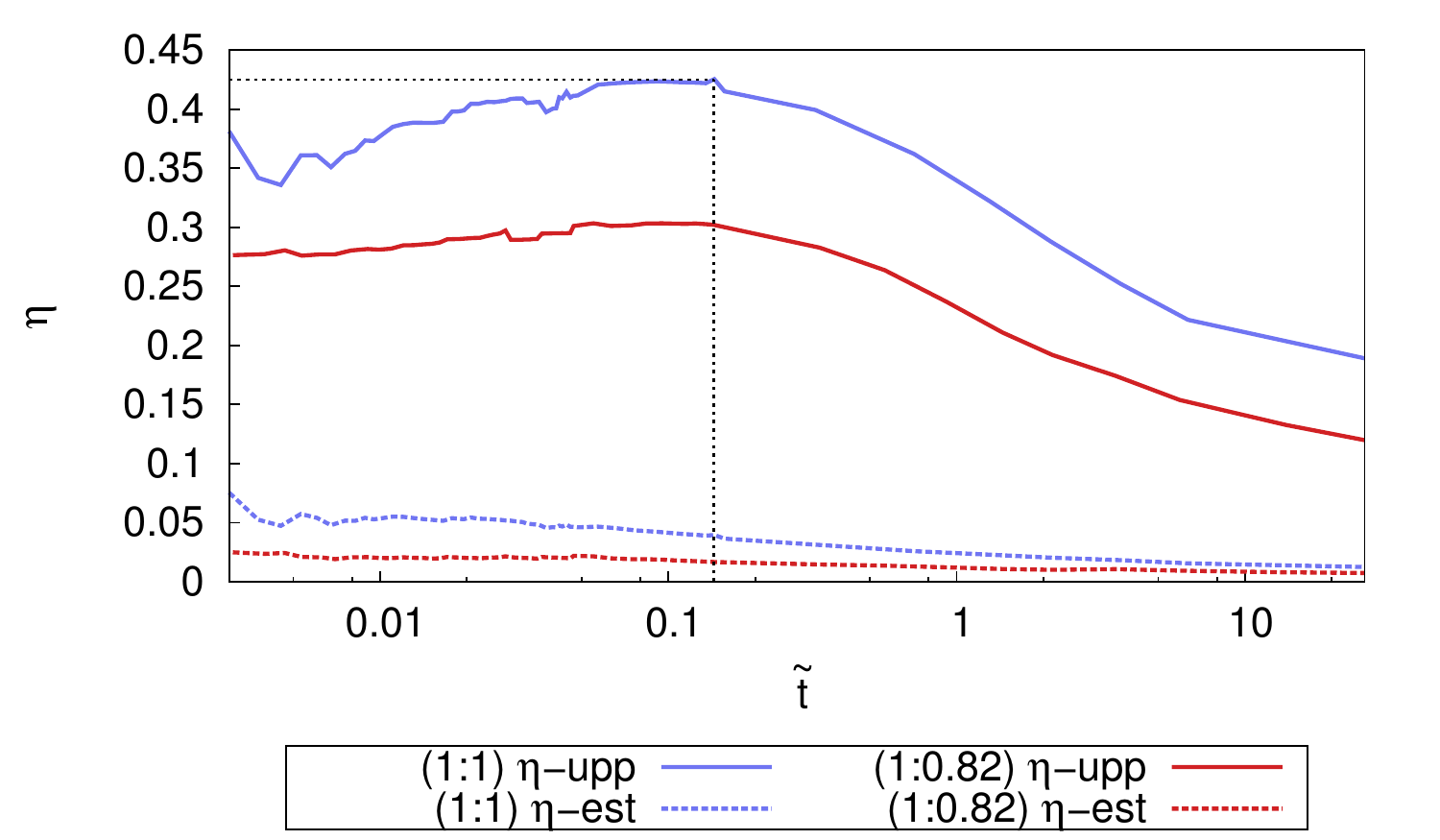}\\
\caption{Efficiencies as a function of annealing time $\tilde{t}$ (dimensionless) for two different blend ratios. Plots show estimates of efficiencies for blend ratios 1:1 (blue) and 1:0.82 (red). (a) Light absorption, (b) exciton dissociation, (c) charge transport, and (d) total efficiency with maximum value marked. }
\label{fig:results:CHeq:const}
\end{figure}

Similar to the previous subsection, we estimate all constitutive efficiencies (for all 1800 morphologies) and analyze them individually.  Figure~\ref{fig:results:CHeq:const} plots the constitutive efficiency estimates. 

We begin with the \emph{light absorption efficiency}. Our first observation is  that values of $\eta_{abs}^{upp}$, the fraction of light absorbing material, does not change with time. This is not surprising, because the mass content is conserved. For blend ratio 1:1, the upper bound of light absorption is $50\%$. While for blend ratio 1:0.82, $\eta_{abs}^{upp}$ is 45\%.\footnote{Small variation is observed in $\eta_{abs}^{upp}$ at early times. This is a discretization artifact.} The same plot also shows estimates of absorption efficiency corrected for the effect of depth $\eta_{abs}^{est}$. 
Estimated trends are similar to the trends for the upper-bounds.
This similarity is a consequence of the following: (i) the height of the film is the same for all morphologies, thus, light penetrates all samples similarly; (ii) for the particular set of morphologies, the substrate is neutral and does not cause  distinct gradients in composition. 

Next, we estimate the \emph{exciton dissociation efficiency}. Thermal annealing increases the feature size. This is clearly seen in Figure~\ref{fig:results:CHeq}. Increasing feature size does not affect exciton dissociation as long as the feature size is smaller than the exciton diffusion length. Beyond this point, fewer excitons reach the donor-acceptor interface, thus decreasing exciton dissociation. We observe this trend for both data sets as seen in Figure~\ref{fig:results:CHeq:const}(b) in which $\eta_{diss}^{upp}$ and $\eta_{diss}^{est}$ are plotted. It is interesting to note that even though the type of morphology is significantly different in both cases, the estimates are comparable, respectively. 

Finally, we compare the two data sets with respect to \emph{charge transport} \emph{efficiency}. In Figure~\ref{fig:results:CHeq:const}(c), we compare two data sets with respect to fraction of useful regions ($\eta_{out}^{upp}$) and fraction of the donor-acceptor interface with complementary paths ($\eta_{out}^{path}$). 
We notice that the coarsening of morphology improves the quality of paths as well as the total fraction of useful domains, as expected. However, unlike estimates of light absorption and exciton dissociation ($\eta_{abs}$ and $\eta_{diss}$), charge transport is clearly affected by the type of morphology.  For blend ratio 1:0.82, $\eta_{out}^{upp}$ is unable to reach 100\%. In contrast, for percolated structure (blend ratio 1:1), it is possible for the entire active layer to be useful and connected to appropriate electrodes. Both estimates $\eta_{out}^{upp}$ and $\eta_{out}^{path}$ are able to discern between classes of morphology. 

Additionally, we notice that for blend ratio 1:0.82 the fraction of useful domain cannot be improved significantly in the course of thermal annealing. In contrast, such improvement is possible for interpenetrated structure (1:1). This is very interesting and shows that not every system can be improved by thermal annealing. This is particularly important in the context of tailoring the fabrication process.

To summarize the analysis, we cumulate all factors and compare the ``total efficiency'' defined as product of all constitutive efficiencies:~\footnote{Several efficiencies of OSC are currently used, e.g., power conversion efficiency, internal quantum efficiency, and external quantum efficiency. The total efficiency we use is different from these but provides important insight into cumulative effect of all constitutive efficiencies.}
\begin{equation}
\eta=\eta_{abs}\cdot\eta_{diss}\cdot\eta_{out}
\end{equation}
\textbf{Remark 4:} The product of $\eta_{diss}$ and $\eta_{out}$ is known as the internal quantum efficiency ($\eta_{IQE}=\eta_{diss}\cdot \eta_{out}$).

In Figure~\ref{fig:results:CHeq:const} (d), we present two plots of the final efficiency estimates: $\eta^{upp}$ and $\eta^{est}$. Estimate $\eta^{upp}$ is constructed using only the upper bounds of efficiencies of constitutive processes: $\eta_{abs}^{upp}$, $\eta_{diss}^{upp}$, $\eta_{out}^{upp}$, 
while $\eta^{est}$ uses the following constitutive estimates: $\eta_{abs}^{est}$, $\eta_{diss}^{est}$, $\eta_{out}^{path}$. 
For both blend ratios, efficiency increases initially, reaches a maximum and subsequently decreases. This trend is due to two competing properties: exciton dissociation and charge transport. The initial increase is because of improved pathways which result in increased charge transport properties. During this period, exciton dissociation remains unaltered and high. 
As morphology evolution proceeds further, exciton dissociation deteriorates rapidly (due to increased feature size). The rate at which exciton dissociation deteriorates is much faster than improvement of charge transport. Consequently, the total efficiency also deteriorates. This shows that during thermal annealing, charge transport is the initial bottleneck, agreeing with a recent work by Chen et al.~\cite{CHGY09}. Beyond a certain point during annealing, exciton dissociation becomes the bottleneck and cannot be compensated by the improving charge transport.

It is also evident that estimated efficiencies are significantly lower for particle-like structures than interpenetrated structures. This emphasizes the importance of morphology control to ensure an interpenetrated structure with minimal number of isolated regions.

From Figure~\ref{fig:results:CHeq:const}, the optimal morphology occurs at $\tilde{t}=0.144$ with constitutive efficiencies of: $\eta_{abs}^{upp}=50\%$, $\eta_{diss}^{upp}=98\%$ and $\eta_{out}^{upp}=86\%$. This translates into an internal quantum efficiency of $\eta_{IQE}=84\%$. This is the internal quantum efficiency reported for the best devices obtained using thermal annealing~\cite{PRB09,LiangYu2010,HouLi2009}.

\section{Conclusion}
\label{ch:C}
A comprehensive set of computational tools to rapidly quantify and classify the 2D/3D heterogeneous internal structure of OSC thin films is invaluable in linking process, structure and property. Development of  such a framework is challenging because of the coupled and contradictory effect of morphology on light absorption, exciton dissociation and charge transport. In this paper we addressed these challenges and provided a highly efficient framework based on graph theory to characterize morphology. In particular: 
\begin{enumerate}
\item we presented the first such broad analysis of the BHJ morphological features, where all subprocesses are quantified and analyzed individually. 
\item we categorized physically meaningful morphology into four main groups.
\item we noticed that all important quantities in this context are based on two types of information: (1)~the distance between two regions of a considered system, and (2)~the connectivity between domains. 
\item we recognized that these two quantities are central elements of graph theory, and also recognized the equivalence between the morphology and labeled, weighted, undirected graph. Both observations motivated us to borrow from graph theory and exploit standard graph algorithms. 
\item we formulated six questions to provide a comprehensive characterization of BHJ using the graph approach. 
\item we additionally employed the questions to provide upper bounds and estimates on all constitutive subprocesses of the photovoltaic effect in organic materials. 
\item we identified the bottlenecks in the morphology evolution during thermal annealing, which is one the major fabrication processes, through the individual analysis of subprocesses, .
\end{enumerate}

The developed framework of morphology descriptors is generic. In particular, the graph approach makes the framework dimension independent. In addition to the 2D results shown in this paper, it is straightforward to utilize the framework for analyzing 3D structures as well as periodic structures~\cite{WTCG11b}. 

A graph-based approach can be naturally extended by adding additional properties to vertices and edges. Specifically, crystallinity effects can be investigated by defining an additional property (degree of crystallinity) with each vertex. Similarly, anisotropy effects can be included by defining direction dependent edge properties. Both extensions are important for full characterization of OSC morphology.

We also envision its application in other areas involving transport in heterogeneous structures, such as percolation pathways in geomechanics, porous media, and drug release from polymeric membranes.

\section{Acknowledgments}
ST was supported in part by NSF-0831903. SC acknowledges financial support from Iowa Power Fund, provided by state of Iowa’s Office of Energy Independence. BG was supported in part by NSF PHY-0941576 and NSF CCF-0917202.

\appendix
\label{App}
\section{Fraction of light absorbing material corrected by absorption depth}
\label{App:LI}
The irradiance of light at a given depth, $x$, of device is given by formula:
\begin{equation}
I=I_0e^{-x/H_d},
\end{equation}
where  $I_0$ is the light irradiance at top surface ($W/m^2$) and $H_d$ is the absorption depth.

The efficiency of light absorption can be determined as a ratio between the total light irradiance penetrating different heights in the device and the light irradiance that would lead to ideal light  absorption by the entire device.
\begin{equation}
\eta_{abs}^{est}=\displaystyle\frac{I_0\int_{0}^{h_{tot}}M(x)e^{-\frac{x}{H_d}}dx}{I_0\int_0^{h_{tot}} dh},
\end{equation}
where $M(x)$ is the binary phase function that accounts for the light absorbing material, it takes value $M(x)=1$ for electron donor region (which is the light absorbing material) and M(x)=0 for electron acceptor region.

\bibliographystyle{model1-num-names}

\end{document}